\documentclass[showpacs,twocolumn,aps,preprintnumbers,letterpaper]{revtex4}
\usepackage{amsmath,amssymb}
\usepackage{epsfig}
\usepackage{graphicx}
\usepackage{amsmath}
\usepackage{slashed}
\usepackage{amsfonts}
\usepackage{epstopdf}
\usepackage{color}
\usepackage[caption=false]{subfig}
\usepackage[pdftex, pdfborder= 0 0 0, citecolor=blue, urlcolor=blue, linkcolor=blue, colorlinks=true, bookmarksopen=true]{hyperref}

\newcommand{\be}{\begin{equation}}
\newcommand{\ee}{\end{equation}}
\newcommand{\bear}{\begin{eqnarray}}
\newcommand{\eear}{\end{eqnarray}}
\newcommand{\ba}{\begin{array}}
\newcommand{\ea}{\end{array}}

\def\be{\begin{eqnarray}}
\def\ee{\end{eqnarray}}
\def\bea{\be}
\def\eea{\ee}

\def\roughly#1{\mathrel{\raise.3ex\hbox{$#1$\kern-.75em%
\lower1ex\hbox{$\sim$}}}}

\def\abs#1{{\left| #1 \right|}}

  \long\def\comment#1{ }

  \newcommand{\Tr}{{\rm Tr}}

  \newcommand{\beq}{\begin{eqnarray}}
  \newcommand{\eeq}{\end{eqnarray}}
  
 \def\simge{\mathrel{%
   \rlap{\raise 0.511ex \hbox{$>$}}{\lower 0.511ex \hbox{$\sim$}}}}
\def\simle{\mathrel{
   \rlap{\raise 0.511ex \hbox{$<$}}{\lower 0.511ex \hbox{$\sim$}}}}

\begin{document}

\title{Electroproduction of heavy vector mesons using holographic QCD:\\ from near threshold to high energy regimes}

\author{Kiminad A. Mamo}
\email{kmamo@anl.gov}
\affiliation{Physics Division, Argonne National Laboratory, Argonne, Illinois 60439, USA
}


\author{Ismail Zahed}
\email{ismail.zahed@stonybrook.edu}
\affiliation{Center for Nuclear Theory, Department of Physics and Astronomy, Stony Brook University, Stony Brook, New York 11794-3800, USA}



\date{\today}
\begin{abstract}
We develop a non-perturbative analysis of the electro-production of heavy vector mesons ($\phi$, $J/\Psi$) from threshold to high energy.
We use the holographic construction with bulk confinement enforced through a soft wall.  Using Witten diagrams,
we evaluate the pertinent cross sections for heavy vector mesons ($\phi$, $J/\Psi$) production and study their dependence on
both the incoming virtual photon polarization as well as the outgoing polarization of the heavy meson. Our results for $J/\Psi$
electro-production compares well with the available HERA data at low and intermediate $Q^2$, and for a wide range of momentum transfer. 
We also predict the quasi-electroproduction of $J/\Psi$ near threshold.
\end{abstract}


\maketitle

\setcounter{footnote}{0}


\section{Introduction}
Diffractive  production of heavy mesons such as charmonia and bottomonia through the  scattering of
real or virtual photons  on a proton, is amenable to the scattering of a virtual hadron on a proton when
the photon coherence length becomes comparable to the nucleon size. By varying the virtuality $Q^2$ 
of the photon and its polarization, a scanning of the reaction as a function of the virtual hadron size and 
light cone content can be carried out.

At small $Q^2$ the photon virtual size is comparable to the hadronic  size, and the diffractive process
is similar to that observed in diffractive hadron-hadron scattering with a strong dependence on the soft
or non-perturbative Pomeron process. With increasing $Q^2$, the virtual size of the photon decreases.
The decrease is sensitive to the virtual polarization and allows for a characterization of the transition from
the non-perturbative  to perturbative physics.

We will address the diffractive problem at low and intermediate $Q^2$ non-perturbatively in the context of holographic QCD, which  embodies
among others the pre-QCD  dual resonance model. The approach originates from a conjecture that observables in 
strongly coupled gauge theories in the limit of a large number of colors, can be determined from classical 
fields interacting through gravity in an anti-de-Sitter space in higher dimensions~\cite{HOLOXX}. The present
study will be a follow up on our recent photo-production analysis of heavy mesons~\cite{Mamo:2019mka}. We
note that exclusive production of heavy mesons in the holographic context has been considered  
in~\cite{DJURIC,LEE,Hatta:2007he,Hatta:2009ra}, and in the non-holographic context in~\cite{MANY}.


Empirical studies of electro-production of heavy mesons have been pioneered at HERA. Data from ZEUS and H1
show that with increasing $Q^2$ the production is enhanced, a point in favor of a transition from a soft Pomeron 
to a hard Pomeron mechanism. The holographic  construction captures this transition through a migration of the low
lying string fluctuations in bulk  from the infrared to the ultraviolet section of the AdS space~\cite{Brower:2006ea,Stoffers:2012zw}. 
For completeness, we note that the holographic  formulation of the Pomeron as a string exchange in bulk, was initiated originally in~\cite{SIN}.

More recently, the GlueX collaboration at Jefferson LAB has turned measurements  of threshold charmonium production
at the photon point~\cite{GLUEX}, which are in the process of further  refined by the  ongoing  measurements from the high precision  $J/\Psi-$007 
collaboration~\cite{MEZIANI}. A chief motivation for these experiments is a measure of the gluonic contribution entering
the composition of the nucleon mass, and possibly the treshold photo-production of the LHCb pentaquark. The importance of the gluon exchange 
in the diffractive production of $J/\Psi$ near treshold has been suggested in~\cite{BRODSKY}, and received 
 considerable attention lately~\cite{Mamo:2019mka,MANYJPSI}.



The organization of the paper is as follows:  In section II we outline the general set up by detailing the pertinent kinematics,
and briefly reviewing the graviton and dilaton bulk actions and couplings essentials for the construction of the diffractive
electro-production amplitude of charmonium using a Witten diagram. In section III, the differential and total cross sections
for $J/\Psi$ electro-production are detailed near treshold and far from treshold for both the transverse and longitudinal
polarizations. Remarkably, in the double limit of large $N_c$ and strong coupling the tensor or A-form factor dominates 
solely the treshold  production, and its Reggeized form its high energy counterpart.
In section IV, the results from near treshold are compared to the GlueX data at the photon point, with the
ensuing predictions for the quasi-electro-production given. In section V, we compare our results for $J/\Psi$ electro-production
to the existing HERA data. We also extend our analysis to the lighter $\phi$-meson production. Our conclusions are in section VI. 
A number of Appendices are added to provide the necessary definitions and details for many of  the sections.

\section{General set up}

In our recent analysis of the holographic photoproduction  of heavy mesons~\cite{Mamo:2019mka}, we noted that even close to treshold
the process was mostly diffractive and dominated by the exchange of a massive tensor
$2^{++}$ graviton at threshold, and higher spin-j exchanges away from threshold that rapidly reggeize.
The scalar $0^{++}$ glueballs were  found to
decouple owing to their vanishing coupling to the virtual photons, while the dilatons were shown to
decouple from the bulk Dirac fermion. At threshold, the holographic photoproduction amplitude solely probes
the gravitational A-form-factor which maps on the gluonic contribution to the energy momentum tensor of 
the nucleon as a Dirac fermion in the bulk. We now extend these observations to the electroproduction  process.

\begin{widetext}

\subsection{Differential cross section and kinematics}

The differential cross section for  electro-production of heavy mesons $V=J/\Psi, \Upsilon$ involves the exchange
of a bulk $j=2$ graviton near threshold, and higher spin away from threshold. We will postpone the higher spin exchanges
and their Reggeization to later. Specifically, consider the DIS process 
$\gamma^*(q,\epsilon)+N(p)\rightarrow V (q^\prime,\epsilon^\prime)+N(p^\prime)$ with 
in $\epsilon$ and out $\epsilon^\prime$ polarizations. The corresponding differential cross section is

\be
\label{DCGENERAL}
\frac{d\sigma(s,t,Q,M_{J/\Psi}, \epsilon, \epsilon^\prime)}{dt}
=\frac{e^2}{16\pi(s-(-Q^2+m_N^2))^2}\,
\frac 12\sum_{{\rm spin}}
\Bigg|{\cal A}_{\gamma^* p\rightarrow  V p} (s,t,Q,M_{J/\Psi},\epsilon,\epsilon^\prime)\Bigg|^2\,,\nonumber\\
\ee
with a virtual and space-like  $q^2=-Q^2<0$, and  Mandelstam  $s=(q+p_1)^2>0$  with $t=(q-q^\prime)^2<0$.




We will analyze (\ref{DCGENERAL}) using holography with mostly negative signature $\eta^{\mu\nu}=(+,-,-,-)$ in 4-dimensions, and in the center-of-mass (CM) frame of the pair composed the virtual photon $\gamma^*$ and the proton. Specifically, for the incoming channel $q=(q_0,0,0,q_z)$ and $p_1=(p_{10},0,0,p_{1z}=-q_z)$
and for the outgoing channel $q^\prime=(q'_0,\mathbf{q}_V)$ and $p_2=(p_{20},-\mathbf{q}_V)$. We also have 
\be
q_z=\vert\mathbf{q}_\gamma\vert=\frac{1}{2\sqrt{s}}\sqrt{s^2-2(-Q^2+m_N^2)s+(-Q^2-m_N^2)^2}\,,\nonumber\\
\ee
\be
\vert\mathbf{q}_V\vert=\frac{1}{2\sqrt{s}}\sqrt{s^2-2(M_V^2+m_N^2)s+(M_V^2-m_N^2)^2}\,,\nonumber\\
\ee
\be
\label{tminmaxx}
t=-Q^2+M_V^2-2E_{\gamma}E_V+2\vert\mathbf{q}_\gamma\vert\vert\mathbf{q}_V\vert\cos\theta\,,\nonumber\\
\ee
$q_0=E_{\gamma}=\sqrt{-Q^2+q_z^2}$, $q_0'=E_{V}=\sqrt{M_V^2+\vert\mathbf{q}_V\vert^2}$, $p_{10}=\sqrt{m_N^2+q_z^2}$, and $p_{20}=\sqrt{m_N^2+\vert\mathbf{q}_V\vert^2}$.

Since $\gamma^*, V$ are treated as gauge particles, we can use the gauge freedom to choose  both polarizations to be 4-transverse to the incoming momenta
or $\epsilon\cdot q=0$ and $\epsilon^\prime\cdot q^\prime=0$. For the incoming $\gamma^*$, we can set the polarizations as

\bea
\epsilon_{T=\pm}=\frac{1}{\sqrt{2}}(0,\mp 1,-i,0)\qquad \epsilon_L=\frac{1}{Q}(q_z,0,0,q_0)
\eea
which satisfy $\epsilon_{T,L}\cdot q=0$ and $\epsilon_{T}\cdot\epsilon_L=0$,  with the real 2-vector normalizations $\epsilon_T^2=-1$ and $\epsilon_L^2=-1$.
For the outgoing $V$, moving in $\mathbf{\hat q}_V=(\sin\theta\cos\phi,\sin\theta\sin\phi,\cos\theta)$ direction, we can set the polarizations (see, for instance, Apendix I.2 of \cite{Dreiner:2008tw})

\bea
\epsilon_{T=\pm}^\prime=\frac{1}{\sqrt{2}}e^{\mp i\gamma}(0,\mp \cos\theta \cos\phi +i\sin\phi,\mp \cos\theta \sin\phi -i\cos\phi,\pm \sin\theta)
\eea
and
\bea
\epsilon^\prime_L=\frac{1}{M_{V}}(\vert\mathbf{q}_V\vert,E_V\sin\theta\cos\phi,E_V\sin\theta\sin\phi,E_V\cos\theta)
\eea
which satisfy $\epsilon^\prime_{T,L}\cdot q^\prime=0$ and $\epsilon^\prime_{T}\cdot\epsilon^\prime_L=0$,  
with the real 2-vector normalizations $\epsilon_T^{\prime 2}=-1$ and $\epsilon_L^{\prime 2}=-1$. 
We also have
\be
\cos\theta=\frac{t+Q^2-M_V^2+2E_{\gamma}E_V}{2\vert\mathbf{q}_\gamma\vert\vert\mathbf{q}_V\vert}\,.
\ee
We can set the phase angle $\gamma=0$ since the differential cross-section is independent of it. We can also use

\begin{figure}[!htb]
\includegraphics[height=5cm]{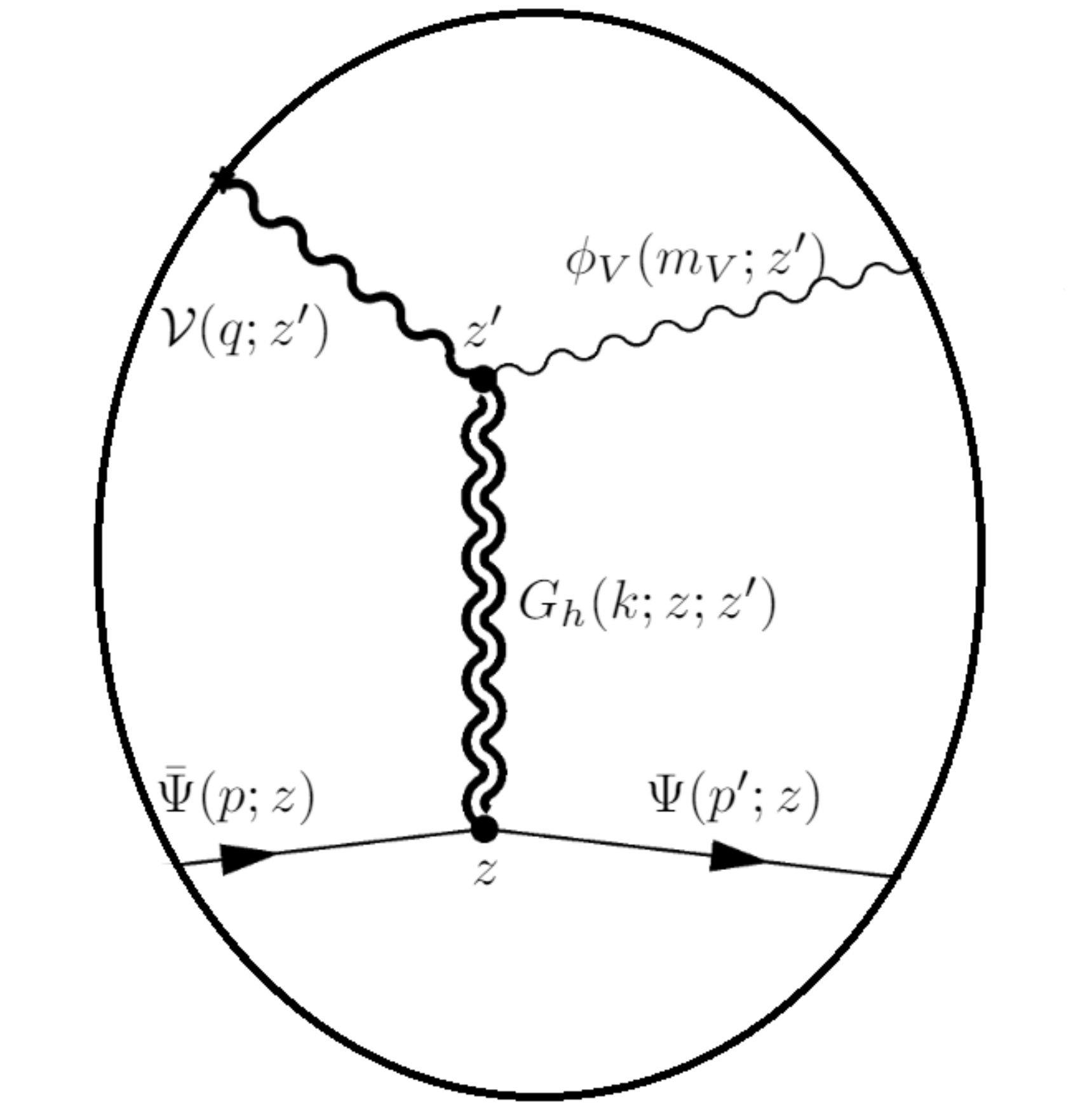}
  \caption{Witten diagram for the diffractive photoproduction of vector mesons with a bulk wave function $\phi_V$. The thick lines or thick wiggles represent the propagators of summed over vector meson or glueball resonances. The thin lines or thin wiggles correspond to a single vector meson and proton. For scalar glueball resonances, due to the dilaton and the trace-full part of the metric fluctuation, we simply replace the bulk-to-bulk propagator $G_h(k,z,z')$ of spin-2 glueballs by $G_{\varphi,f}(k,z,z')$.}
  \label{wdiagram}
\end{figure}

\subsection{Graviton and dilaton bulk action}

Diffractive electro-production on a bulk Dirac fermion in holography, involves the exchange of tensor and scalar gravitons.
The tensor graviton exchange is dual to a Reggeized tensor $2^{++}$ glueball, and the scalar graviton exchange is dual to
a Reggeized scalar $0^{++}$ glueball. This is  illustrated in the Witten diagram of  Fig.~\ref{wdiagram}.
The generic effective action  for these diagrams  in  AdS$_5$ is given by

\be
{\cal S}=\int d^5x\,e^{-2\phi(z)}\sqrt{g}\,{\cal L}_{EH}(g, \phi)   +\int d^5x\,e^{-\phi(z)}\sqrt{g}\,{\cal L}_{DBI}(\psi,A,X) 
\ee
with the Einstein-Hilbert action (EH) in the string frame for the metric $g$, $\phi$-dilaton,  
and the DBI action for the Dirac fermion  $\psi$, flavor gauge field $A$, and tachyon $X$. The additional dimensionally reduced 
string modes,   not entering the present discussion,  have been omitted. The AdS$_5$ metric is chosen as
$g_{MN}=(\eta_{\mu\nu}, -1)/z^2$ with $\eta_{\mu\nu}$ mostly negative.

The fluctuations in the 4-dimensional part of the bulk metric, 
split into a transverse and traceless TT-part denoted by $h$ (tensor  graviton) and a transverse and traceful T-part 
denoted by $f$ (scalar  glueball)

\be
\label{HMUNU}
g_{\mu\nu}(z)\rightarrow g_{\mu\nu}(z)+ \sqrt{2\kappa^2}\bigg[h_{\mu\nu}(x,z)\approx \epsilon_{\mu\nu}^{TT}\,h(x,z)+\frac 14\eta_{\mu\nu}\,f(x,z)\bigg]
\ee
with $k^\mu\epsilon_{\mu\nu}^{TT}=\eta^{\mu\nu}\epsilon_{\mu\nu}^{TT}=0$. The Newtonian coupling is fixed by the 
D-brane tension with $16\pi G_N=2\kappa^2=8\pi^2/N_c^2$.
The decomposition (\ref{HMUNU}) follows in the gauge where $f,h$ decouple. However, they both obey identical  equations of motion 
since they carry the same anomalous dimensions $\Delta_{T,S}=4$ in the strict large $N_c$ limit~\cite{Kanitscheider:2008kd}. This is
not the case at finite $1/N_c$ (which will be subsumed below). 
More specifically, the radial Regge spectrum of tensor and scalar glueballs, is~\cite{HILMAR}

\be
m_{T,S}^2=8\kappa_N^2\bigg(n+2\bigg)
\ee
for a background dilaton $\phi(z)=\kappa_N z^2$.

\subsection{Graviton and dilaton bulk couplings}

The tensor graviton and scalar dilaton coupling to the energy-momentum of the $\frac 12$ Dirac fermion and $1^{--}$ flavor vector fields
in bulk is given by~\cite{Mamo:2019mka}
 
  \be
-\frac{\sqrt{2\kappa^2}}{2}\int d^5x\,\sqrt{g}\,h_{\mu\nu}(T_F^{\mu\nu}+T_V^{\mu\nu})=
-\frac{\sqrt{2\kappa^2}}{2}\int d^5x\,\sqrt{g}\,\bigg(\epsilon_{\mu\nu}^{TT}\,h+\tilde{k}^2\frac 14\eta_{\mu\nu}\,f\bigg)\,(T_F^{\mu\nu}+T_V^{\mu\nu})
\label{vertices1}
 \ee
where the energy-momentum tensors are

 \bea
 \label{TMUNU}
T_F^{\mu\nu}&=&e^{-\phi_N}\frac{i}{2}\,z\,\overline\Psi\gamma^\mu\overset{\leftrightarrow}{\partial^\nu}\Psi-\eta^{\mu\nu}\mathcal{L}_F\,,\nonumber\\
T_V^{\mu\nu} &=&-e^{-\phi_V}\Big(z^4\eta^{\rho\sigma}\eta^{\mu\beta}\eta^{\nu\gamma}\,F^V_{\beta\rho}F^V_{\gamma\sigma}
-z^4\,\eta^{\mu\beta}\eta^{\nu\gamma}\,F^V_{\beta z}F^V_{\gamma z}\Big)-\eta^{\mu\nu}\mathcal{L}_V\,,
  \label{EMT}
 \eea
with $\phi_N=\tilde{\kappa}_{N}^2z^2$ for the nucleon field,
 and $\phi_V=\tilde{\kappa}_{V}^2z^2$ for the  heavy mesons $\bar s\gamma^\mu s, \bar c\gamma^\mu c, \bar b\gamma^\mu b$  or $\phi,J/\Psi,\Upsilon$,
 respectively. The photon field follows from a similar reasoning with $\kappa_V=\kappa_\rho$.
 All vector fields are massless in bulk, since they are $p=1$ forms with anomalous dimension $\Delta_V=3$, 
 
 \be
 m^2_5=(\Delta-p)(\Delta +p-4)\rightarrow 0
 \ee
Their gauge-coupling to a background tachyon field $X\sim q\bar q$ in bulk with  a massive boundary condition
$X_0\sim {\rm diag}(m_u,m_d,m_c, m_b)$, only generates a Higgs mass for the off-diagonal or charged
flavor fields, the diagonal or neutral fields dual to  $\phi, J/\Psi,\Upsilon$ remain massless in bulk. A mass $m_{c,b}$
can be added by minimally modifying the dilaton potential $\phi(z)$, without affecting the traceless condition
of $T_V^{\mu\nu}$.

\subsection{Diffractive electro-production amplitude}

The scalar dilaton coupling to the flavor vector fields vanishes in bulk, i.e.
 $\eta_{\mu\nu}T_V^{\mu\nu}=0$  in (\ref{TMUNU}), and drops out  of 
 the diffractive process in~ Fig.~\ref{wdiagram}~\cite{Mamo:2019mka},
 for both the real and virtual incoming photons.
 As a result, the electro-production amplitude follows solely from the exchange of 
 a tensor glueball or graviton,


\be
\label{nAmph}
i{\cal A}^{h}_{\gamma^* p\rightarrow  J/\Psi p} (s,t)\approx \frac{1}{g_5}\times(-i)\mathcal{V}^{\mu\nu}_{h\gamma^* J/\Psi}(q, q^\prime, ,k_z)\times \bigg(\frac{i}{2}\eta_{\mu\alpha}\eta_{\nu\beta}\bigg)\times(-i)\mathcal{V}^{\alpha\beta}_{h\bar\Psi\Psi}(p_1,p_2,k_z)\,,
\ee
with $k=p_2-p_1=q-q^\prime$ and 

\be
&&\mathcal{V}^{\mu\nu}_{h\gamma^* J/\Psi}(q,q^\prime,k_z)=\sqrt{2\kappa^2}\times\frac{1}{2}\int dz\sqrt{g}\,e^{-\phi_{J/\psi}}z^4K^{\mu\nu}(q,q',\epsilon,\epsilon',z)\frac{z^4}{4}\,,\nonumber\\
&&\mathcal{V}_{h\bar\Psi\Psi}^{\alpha\beta}(p_1, p_2,K)
=-\sqrt{2\kappa^2}\times\frac{1}{2}\int dz\sqrt{g}\,e^{-\phi_N}z\,\big(\psi_R^2(z)+\psi_L^2(z)\big)\mathcal{H}(K,z)\times\bar u(p_2)\gamma^\alpha p^\beta u(p_1)\,,
\ee
We  have set  $p=(p_1+p_2)/2$, $q^2=-Q^2$ and $q^{\prime 2}=-Q^{\prime 2}$  for space-like momenta. The tensors are defined as

\bea
K^{\mu\nu}(q,q',\epsilon,\epsilon',z)\equiv && B_1^{\mu\nu}\mathcal{V}_{\gamma^*}(Q,z)\mathcal{V}_{J/\Psi}(M_{J/\Psi},z)
-B_0^{\mu\nu}\partial_z\mathcal{V}_{\gamma^*}(Q,z)\partial_z\mathcal{V}_{J/\Psi}(M_{J/\Psi},z)\,,\\
\nonumber\\
B_{0}^{\mu\nu}(\epsilon,\epsilon')\equiv &&\epsilon^\mu \epsilon'^\nu\,,\nonumber\\
B_{1}^{\mu\nu}(q,q',\epsilon,\epsilon')\equiv&&
\epsilon\cdot \epsilon' \,q^\mu q'^\nu-q\cdot \epsilon'\, \epsilon^\mu q'^\nu
-q^\prime \cdot \epsilon\, q^\mu \epsilon^{\prime \nu}+q\cdot q^\prime\, \epsilon^\mu \epsilon^{\prime \nu}\,.\\\nonumber
\label{BK}
\eea
with $B_{1,0}=\eta_{\mu\nu}B_{1,0}^{\mu\nu}$, and $K=\eta_{\mu\nu}K^{\mu\nu}$.

The non-normalizable wave function for the virtual photon $V_{\mu}(Q,z)=\mathcal{V}_{\gamma^*}(Q,z)\,\epsilon_{\mu}e^{-iq\cdot x}$ 
is the bulk-to-boundary propagator for the Reggeized process $\gamma^*\rightarrow c\bar c$

\be
\label{BTOBV}
\mathcal{V}_{\gamma^*}(Q,z)
=\tilde{\kappa}_{J/\Psi}^2z^2\int_{0}^{1}\frac{dx}{(1-x)^2}x^a{\rm exp}\Big[-\frac{x}{1-x}\tilde{\kappa}_{J/\Psi}^2z^2\Big]\,,
\ee
with  $a={Q^2}/{4\tilde{\kappa}_N^2}$ and the normalization ${\cal V}_{\gamma^*}(0,z)={\cal V}_{\gamma^*}(Q,0)=1$. 
The normalizable wave function for $J/\Psi$ is $V_{\mu}(q',z)=\mathcal{V}_{J/\Psi}(M_{J/\Psi},z)\,\epsilon'_{\mu}e^{-iq'\cdot x}$ where 

\be
\label{JPSI}
\mathcal{V}_{J/\Psi}(M_{J/\Psi},z)
=\phi_{0}(M_{J/\Psi},z)=\frac{f_{J/\Psi}}{M_{J/\Psi}}\times 2g_{5}\tilde{\kappa}_{J/\Psi}^2z^2 L_0^1( \tilde{\kappa}_{J/\Psi}^2z^2)\,.
\ee

The non-normalizable wave function for the virtual tansverse and traceless graviton is   
given by $h_{\mu\nu}(Q,z)=\mathcal{H}(Q,z)\epsilon_{\mu\nu}^{TT}e^{-iq\cdot x}$~\cite{Hong:2004sa,CARLSON,Abidin:2008ku,BallonBayona:2007qr}

\be
\label{BTOBH}
\mathcal{H}(Q,z)=&&4z^{4}\Gamma(a_Q+2)U\Big(a_Q+2,3;2\tilde{\kappa}_N^2z^2\Big) =\Gamma(a_Q+2)U\Big(a_Q,-1;2\tilde{\kappa}_N^2z^2\Big)\nonumber\\
=&&\frac{\Gamma(a_Q+2)}{\Gamma(a_Q)} \int_{0}^{1}dx\,x^{a_Q-1}(1-x){\rm exp}\Big(-\frac{x}{1-x}(2\tilde{\kappa}_N^2z^2)\Big)
\eea
with $a_Q={Q^2}/{8\tilde{\kappa}_N^2}$, and we have used the transformation $U(m,n;y)=y^{1-n}U(1+m-n,2-n,y)$. 
(\ref{BTOBH}) satisfies the normalization condition ${\cal H}(0,z)={\cal H}(Q,0)=1$.


\section{Differential and total cross sections for electroproduction}


\subsection{Near threshold}

The differential cross section for  electro-production of heavy mesons $J/\Psi$ (and also $\Upsilon$) involves the exchange
of a bulk $j=2$ graviton near threshold, and higher spin away from threshold. We will postpone the higher spin exchanges
and their Reggeization to later. 

\subsubsection{$TT$ and $LL$ differential cross sections near threshold}

The differential cross section  for untraced in and out polarizations, near threshold, is given by (see the detailed derivations in Appendix \ref{dcnt})
\bea
\label{DCTTLL}
\frac{d\sigma(s,t,Q,M_{J/\Psi},\epsilon_{T},\epsilon'_{T})}{dt}
&=&\mathcal{I}^2(Q,M_{J/\Psi})\times\left(\frac{s}{\tilde{\kappa}_N^2}\right)^{2}\times\mathcal{N}^{TT}(s,t,Q,M_{J/\Psi},m_N)\times\left(-\frac{t}{4m_N^2}+1\right)\times \tilde{A}^2(t)\,,\nonumber\\
\frac{d\sigma(s,t,Q,M_{J/\Psi},\epsilon_{L},\epsilon'_{L})}{dt}
&=&\mathcal{I}^2(Q,M_{J/\Psi})\times\left(\frac{s}{\tilde{\kappa}_N^2}\right)^{2}\times\frac{1}{9}\times\frac{Q^2}{M_{J/\Psi}^2}\times\mathcal{N}^{TT}(s,t,Q,M_{J/\Psi},m_N)\times\left(-\frac{t}{4m_N^2}+1\right)\times \tilde{A}^2(t)\,,\nonumber\\
\eea
where 
\bea
 \label{IJJ}
\mathcal{I}(Q,M_{J/\Psi})&=&\frac{3}{2}\frac{g_5f_{J/\Psi}}{M_{J/\Psi}}\times\frac{1}{\left(\frac{Q^2}{4 \tilde{\kappa}_{J/\Psi}^2}+3\right)\left(\frac{Q^2}{4 \tilde{\kappa}_{J/\Psi}^2}+2\right)\left(\frac{Q^2}{4 \tilde{\kappa}_{J/\Psi}^2}+1\right)}\nonumber\\
&=&\frac{3}{2}\frac{g_5f_{J/\Psi}}{M_{J/\Psi}}\times\tilde{\mathcal{I}}(Q,\tilde{\kappa}_{J/\Psi})\,,
\eea
with $\tilde\kappa_{J/\Psi}=\sqrt{g_5f_{J/\Psi}M_{J/\Psi}}/2^{3/4}$ or $g_5=\frac{2^{3/2}\tilde\kappa_{J/\Psi}^2}{f_{J/\Psi}M_{J/\Psi}}$, and we have defined
\be\label{IQth}
\tilde{\mathcal{I}}(Q,\tilde{\kappa}_{J/\Psi})=\frac{1}{\left(\frac{Q^2}{4 \tilde{\kappa}_{J/\Psi}^2}+3\right)\left(\frac{Q^2}{4 \tilde{\kappa}_{J/\Psi}^2}+2\right)\left(\frac{Q^2}{4 \tilde{\kappa}_{J/\Psi}^2}+1\right)}\,.
\ee

Note that (\ref{IJJ}) follows from
the boundary-to-bulk vector propagator that resums the $1^{--}$  $c\bar c$ radial Regge trajectory, in short the  $1^{--}$
transition form factor   $[\gamma^* \rightarrow c\bar c]+{\rm graviton }\rightarrow J/\Psi$.

The tensor form factor $A(t)$ with  $t=-Q^2<0$, corresponds to the elastic vertex ${\rm graviton }+p\rightarrow p$. It is composed of 
the bulk-to-bulk graviton propagator which resums the
$2^{++}$ radial Regge trajectory (\ref{REGGERHON}),

\bea \label{FFj22}
\tilde{A}(t)=\frac{A(t)}{A(0)}&=&\left(a_K+1\right) \left(2 \left(2 a_K^3+a_K\right) \Phi \left(-1,1,a_K\right)-\left(2 a_K^2+a_K+1\right)\right)\nonumber\\
&=&6\times\frac{\Gamma \left(2+a_K\right)}{\Gamma \left(4+a_K\right)}\times \, _2F_1\left(3,a_K;a_K+4;-1\right)\nonumber\\
&=& \bigg((1-2a_K)(1+a_K^2)+a_K(1+a_K)(1+2a_K^2)\bigg(H\bigg(\frac{1+a_K}{2} \bigg)-H\bigg(\frac{a_K}{2}\bigg)\bigg)\bigg)
\eea
with $a_K={-t}/{8\tilde\kappa_N^2}$.
$H(x)=\psi(1+ x)+\gamma$ is the harmonic number or the di-gamma function plus Euler number.  We fix 
 $\tilde\kappa_N=0.350$ GeV to reproduce both the nucleon ${\frac 12}^+$ and the $1^{--}$ rho radial Regge trajectories

 \be
 \label{REGGERHON}
 m_N^2(n)=4\tilde\kappa^2_N(n+\tau_N-1)=4\tilde\kappa^2_N(n+2)\qquad\qquad m_\rho^2(n)=4\tilde\kappa^2_N\bigg(n+\frac {\Delta_\rho -p}2\bigg)=4\tilde\kappa^2_N(n+1)
 \ee
 with $m_N(0)\approx 990$ MeV and $m_\rho(0)\approx 700$ MeV, and $\tau_N=3$ to reproduce the hard scattering rules.  With this in mind, (\ref{FFj22}) is well parametrized by a dipole
 
 \be
\tilde{A}(t)=\frac{A(t)}{A(0)}\approx  \frac{1}{\left(1-\frac{t}{1.124^2}\right)^2}\,.
 \ee
From (\ref{FFj22}) the mass radius of the proton from the exchange of a tensor glueball is

\be
\langle r^2_{Mass}\rangle =6\bigg(\frac{d{\rm ln} A(t)}{dt}\bigg)_0=\frac {1.04}{{\tilde\kappa}_N^2}=(0.57\,{\rm fm})^2
\ee
A slightly improved assessment of the mass radius can be achieved by  fixing $\tilde\kappa_N$ using the physical rho mass in (\ref{REGGERHON}),
and a smaller  nucleon twist factor $\tau_N=3\rightarrow 2.465$ by using the nucleon mass as in our  recent charge radii re-analyses in~\cite{Mamo:2021jhj}. This will not be pursued
here, to keep the present electro-production analysis in line with our photon-production analysis in~\cite{Mamo:2019mka} for comparison.

The transverse $TT$ and longitudinal $LL$ normalizations, for heavy mesons, are defined as (assuming $s\sim s_{th}\gg m_{N}^2, Q^2, M_{V}^2$, and $|t|\ll s_{th}$ in (\ref{NTTLL}))

\bea\label{NTTLL2}
\mathcal{N}^{TT}(s,t,Q,M_{J/\Psi},m_N)&=&\frac{e^2\times\frac{(2\kappa^2)^2}{g_5^2}}{16\pi}\,
\frac 12\times\frac{\tilde{\kappa}_N^4}{\tilde{\kappa}_{J/\Psi}^8}\times A^2(0)\times\frac{1}{16}\times 8=constant\,,\nonumber\\
\mathcal{N}^{LL}(s,t,Q,M_{J/\Psi},m_N)&=&\mathcal{N}^{TT}(s,t,Q,M_{J/\Psi},m_N)\,,\nonumber\\
\eea
where we used $\frac{1}{s^4}\tilde{F}(s)=\frac{1}{16}$ in (\ref{NTTLL}).



\subsubsection{$TT+LL$  differential cross section near threshold}

The total differential cross section 

\be\label{dctotalth}
\left(\frac{d\sigma}{dt}\right)_{{\rm tot}}=\frac{d\sigma(s,t,Q,M_{J/\Psi},\epsilon_{T},\epsilon'_{T})}{dt}+\frac{d\sigma(s,t,Q,M_{J/\Psi},\epsilon_{L},\epsilon'_{L})}{dt}\,,
\ee
is the sum of the transverse and longitudinal contributions, which takes the explicit form

\be\label{newdctotalth}
\left(\frac{d\sigma}{dt}\right)_{{\rm tot}}=\mathcal{N}_t(s,Q;{\mathbb{\tilde{N}}}^{\prime})\times\Big(-\frac{t^2}{4m_N^2}+1\Big)\times\tilde{A}^2(t)\,, 
\ee
The overall normalization in (\ref{newdctotalth})

\bea\label{normdctotalth}
\mathcal{N}_t(s,Q;\mathbb{\tilde{N}}^\prime)&\equiv& \mathcal{I}^2(Q,M_{J/\Psi})\times\left(\frac{s}{\tilde{\kappa}_N^2}\right)^{2}\times\mathcal{N}^{TT}(s,t,Q,M_{J/\Psi},m_N)\nonumber\\
&\times& \left(1+\mathcal{\tilde{N}}_{R}^2\times\frac{1}{9}\times\frac{Q^2}{M_{J/\Psi}^2}\right)\times \mathbb{\tilde{N}}^\prime\,,
\eea
is fixed by our preceding arguments. Strict bulk-to-boundary correspondence implies $\mathbb{\tilde{N}}^\prime=1$ and $\mathcal{\tilde{N}}_{R}=1$ in the double limit of large $N_c$ and strong gauge coupling 
$\lambda$. Here we assume proportionality between the bulk and boundary with $\mathbb{\tilde{N}}^\prime$ and $\mathcal{\tilde{N}}_{R}$ overall parameters that captures the finite $N_c$ corrections. They will be fixed 
by the best fit to the data below.

\begin{figure}[!htb]
\includegraphics[height=5.5cm]{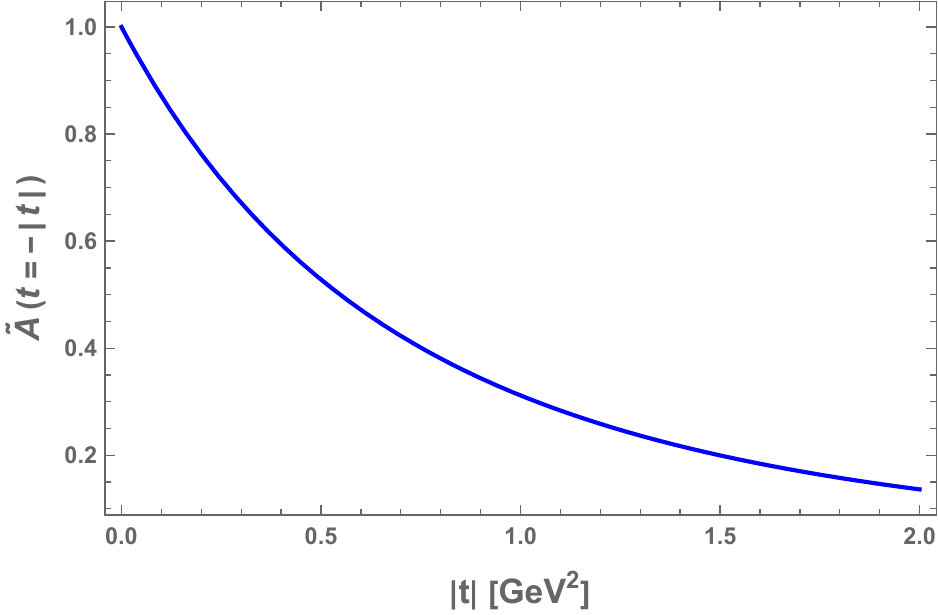}
  \caption{The spin-2 nucleon gravitational form factor $\tilde{A}(t=-|t|)$ given in (\ref{FFj22}).}
  \label{fig_AJT}
\end{figure}

\subsubsection{Total cross section near threshold}

The total cross section for electro-production of a vector meson follows from the differential cross section  by integrating the differential cross section from $t_{min}$ to $t_{max}$, i.e.,
\be\label{totalcsth}
\sigma_{tot}(s,Q^2)=\int_{t_{min}}^{t_{max}}\,dt\left (\frac{d\sigma}{dt}\right)_{{\rm tot}}\,,
\ee
where 
\be\label{dctotalth}
\left(\frac{d\sigma}{dt}\right)_{{\rm tot}}=\frac{d\sigma(s,t,Q,M_{J/\Psi},\epsilon_{T},\epsilon'_{T})}{dt}+\frac{d\sigma(s,t,Q,M_{J/\Psi},\epsilon_{L},\epsilon'_{L})}{dt}\,,
\ee
and, $t_{min}\equiv |t\vert_{\cos\theta=+1}|$ and $t_{max}\equiv|t\vert_{\cos\theta=-1}|$ with $t$ given by (\ref{tminmax}), see also Fig.~\ref{tmx}.

We can also define 
\be\label{sigmaTth}
\sigma_{T}(s,Q^2)=\int_{t_{min}}^{t_{max}}\,dt\frac{d\sigma(s,t,Q,M_{J/\Psi},\epsilon_{T},\epsilon'_{T})}{dt}\,,
\ee
and 
\be\label{sigmaLth}
\sigma_{L}(s,Q^2)=\int_{t_{min}}^{t_{max}}\,dt\frac{d\sigma(s,t,Q,M_{J/\Psi},\epsilon_{L},\epsilon'_{L})}{dt}\,.
\ee
A more  explicit form of the total cross section  (\ref{totalcsth}) is 


\be\label{newtotalcsth}
\sigma_{tot}(s,Q^2)=\mathcal{N}_{Q^2}(s,f_{J/\Psi},M_{J/\Psi};\mathbb{\tilde{N}})\times\mathcal{\tilde{I}}(Q,\tilde{\kappa}_{J/\Psi})\times\left(1+\mathcal{\tilde{N}}_{R}^2\times\frac{1}{9}\times\frac{Q^2}{M_{J/\Psi}^2}\right)\,,\nonumber\\ 
\ee
where $\mathcal{\tilde{I}}(Q,\tilde{\kappa}_{J/\Psi})$ is given by (\ref{IQth}), and we have defined

\bea
\mathcal{N}_{Q^2}(s,f_{J/\Psi},M_{J/\Psi};\mathbb{\tilde{N}})\equiv
\mathcal{N}^{TT}(s,t,Q,M_{J/\Psi},m_N)\times\frac{3}{2}\frac{g_5f_{J/\Psi}}{M_{J/\Psi}}\times 
\mathbb{\tilde{N}}\times\left(\frac{s}{\tilde{\kappa}_N^2}\right)^{2}\times \int_{t_{min}}^{t_{max}}d|t|\,
\left(\frac{|t|}{4m_N^2}+1\right)\times \tilde{A}^2(t=-|t|)\,.\nonumber\\
\eea
Again, we  assume proportionality between the bulk and boundary with $\mathbb{\tilde{N}}$ and $\mathcal{\tilde{N}}_{R}$ overall parameters that capture the finite $N_c$ corrections. They will be fixed by the best fit to the data below. Also note that $\mathcal{N}^{TT}(s,t,Q,M_{J/\Psi},m_N)$ is given by (\ref{NTTLL2}), and it is a constant independent of $Q$, $t$ and $s$.

\subsection{Far from threshold}

The differential cross sections (\ref{DCTTLL}) grow as $s^2$  following the exchange of spin $j=2$ as a tensor glueball 
in bulk. At larger $\sqrt{s}$ higher spin-j exchanges contribute. Their resummation Reggeizes leading to a soft Pomeron
exchange. This transmutation from a graviton to a Pomeron was initially  discussed in~\cite{Brower:2006ea}. In this
section we apply this resummation to the electro-production process.

\subsubsection{$TT$ and $LL$ differential cross sections far from threshold}

More specifically, the differential cross section  for the  untraced in and out polarizations, in the high energy regime, is given by (see the detailed derivations in Appendix \ref{dcher})
\bea
\label{DCTTLLher}
\frac{d\sigma(s,t,Q,M_{J/\Psi},\epsilon_{T},\epsilon'_{T})}{dt}
&=&\mathcal{I}^2(j_0,Q,M_{J/\Psi})\times\left(\frac{s}{\tilde{\kappa}_N^2}\right)^{2\left(1-\frac{2}{\sqrt\lambda}\right)}\times\mathcal{N}^{TT}(j_0,s,t,Q,M_{J/\Psi},m_N)\times \mathcal{A}^2(j_0,\tau,\Delta,t)\,,\nonumber\\
\frac{d\sigma(s,t,Q,M_{J/\Psi},\epsilon_{L},\epsilon'_{L})}{dt}
&=&\mathcal{I}^2(j_0,Q,M_{J/\Psi})\times\left(\frac{s}{\tilde{\kappa}_N^2}\right)^{2\left(1-\frac{2}{\sqrt\lambda}\right)}\times\mathcal{N}^{LL}(j_0,s,t,Q,M_{J/\Psi},m_N)\times \mathcal{A}^2(j_0,\tau, \Delta,t)\,.\nonumber\\
\eea
The $TT$ and $LL$ normalizations in (\ref{DCTTLLher}) are purely kinematical in origin

\bea
\mathcal{N}^{TT}(j_0,s,t,Q,M_{J/\Psi},m_N)&=&\frac{e^2\times\frac{(2\kappa^2)^2}{g_5^2}}{16\pi}\,
\frac 12\times P(\tilde{s},\lambda)\times\frac{\tilde{\kappa}_N^4}{\tilde{\kappa}_{J/\Psi}^8}\times\Bigg(\frac{\tilde{\kappa}_N^2}{\tilde{\kappa}_{J/\Psi}^2}\Bigg)^{-2\big(1+\frac{1}{\sqrt{\lambda}}\big)}\times A^2(0)\times\frac{1}{s^4}\tilde{F}(s)\times 8\nonumber\\
\mathcal{N}^{LL}(j_0,s,t,Q,M_{J/\Psi},m_N)&=&\left(\frac{1}{2-\frac{1}{\sqrt{\lambda}}}\right)^2\times\frac{Q^2}{M_{J/\Psi}^2}\times\mathcal{N}^{TT}(j_0,s,t,Q,M_{J/\Psi},m_N)\,,
\eea
with 

\be
\tilde{F}(s)=\frac{1}{16}\times s^4\,,
\ee
and
\bea \label{P2}
P(\tilde{s},\lambda)&\equiv & \left[\lambda/\pi^2+ 1\right] ( \sqrt{\lambda}/ 2 \pi )\; \tilde{\xi}^2  \; \frac{e^ {-2\sqrt\lambda  \tilde{\xi}^2 / 2\tilde{\tau}}}{\tilde{\tau}^{3}}\left(1 + {\cal O}\bigg(\frac{\sqrt{\lambda}}{\tilde{\tau}}\bigg) \right)\,,
\eea
where $\tilde{\tau}\equiv\log\tilde{s}=\log[s/\tilde{\kappa}_N^2]$ and $\tilde{\xi}-\pi/2=\gamma=0.55772.....$ is Euler-Mascheroni constant.

The transition form factor for $\gamma^*+\mathbb P\rightarrow J/\Psi$ is

\bea
 \label{IJJher2}
\mathcal{I}(j_0,Q,M_{J/\Psi})
&=&\frac{1}{2}\frac{g_5f_{J/\Psi}}{M_{J/\Psi}}\times\frac{\Gamma \left(\frac{Q^2}{4 \tilde{\kappa}_{J/\Psi}^2}+1\right)}{\Gamma\left(\frac{Q^2}{4\tilde{\kappa}_{J/\Psi}^2}+1-\frac{1}{\sqrt{\lambda}}\right)}\times\left(2-\frac{1}{\sqrt{\lambda}}\right)\times\frac{1}{4}\Gamma^2 \left(2-\frac{1}{\sqrt{\lambda}}\right)\nonumber\\
&\times &\frac{1}{\left(\frac{Q^2}{4\tilde{\kappa}_{J/\Psi}^2}+2-\frac{1}{\sqrt{\lambda}}\right)\left(\frac{Q^2}{4\tilde{\kappa}_{J/\Psi}^2}+1-\frac{1}{\sqrt{\lambda}}\right)}
\,,\nonumber\\
&=&\frac{1}{2}\frac{g_5f_{J/\Psi}}{M_{J/\Psi}}\times\left(2-\frac{1}{\sqrt{\lambda}}\right)\times\frac{1}{4}\Gamma^2 \left(2-\frac{1}{\sqrt{\lambda}}\right)\times \mathcal{\tilde{I}}(\lambda,Q,\tilde{\kappa}_{J/\Psi})\nonumber\\
\eea
with 
\be\label{IQ}
\mathcal{\tilde{I}}(\lambda,Q,\tilde{\kappa}_{J/\Psi})\equiv\frac{\Gamma \left(\frac{Q^2}{4 \tilde{\kappa}_{J/\Psi}^2}+1\right)/\Gamma\left(\frac{Q^2}{4\tilde{\kappa}_{J/\Psi}^2}-\frac{1}{\sqrt{\lambda}}\right)}{\left(\frac{Q^2}{4\tilde{\kappa}_{J/\Psi}^2}-\frac{1}{\sqrt{\lambda}}\right)\left(\frac{Q^2}{4\tilde{\kappa}_{J/\Psi}^2}+2-\frac{1}{\sqrt{\lambda}}\right)\left(\frac{Q^2}{4\tilde{\kappa}_{J/\Psi}^2}+1-\frac{1}{\sqrt{\lambda}}\right)}\,,
\ee

with $\tilde\kappa_{J/\Psi}=2^{-3/4}\sqrt{g_5f_{J/\Psi}M_{J/\Psi}}$, which Reggeizes the  $1^{--}$ trajectory, with a substantial fall-off at large $Q^2$.

\subsubsection{$TT+LL$  differential cross section far from threshold}

The total differential cross section 

\be\label{dctotal}
\left(\frac{d\sigma}{dt}\right)_{{\rm tot}}=\frac{d\sigma(s,t,Q,M_{J/\Psi},\epsilon_{T},\epsilon'_{T})}{dt}+\frac{d\sigma(s,t,Q,M_{J/\Psi},\epsilon_{L},\epsilon'_{L})}{dt}\,,
\ee
is the sum of the transverse and longitudinal contributions, which takes the explicit form

\be\label{newdctotal}
\left(\frac{d\sigma}{dt}\right)_{{\rm tot}}=\mathcal{N}_t(s,Q;{\mathbb N}^{\prime})\times\Big(-\frac{t^2}{4m_N^2}+1\Big)\times\mathcal{A}^2(j_0,\tau,\Delta,t)\,, 
\ee
The overall normalization in (\ref{newdctotal})

\bea
\mathcal{N}_t(s,Q;\mathbb N^\prime)&\equiv& \mathcal{I}^2(j_0,Q,M_{J/\Psi})\times\left(\frac{s}{\tilde{\kappa}_N^2}\right)^{2-\frac{4}{\sqrt{\lambda}}}\times\mathcal{N}^{TT}(j_0,s,t,Q,M_{J/\Psi},m_N)\nonumber\\
&\times& \left(1+\mathcal{N}_{R}^2\times\left(\frac{1}{2-\frac{1}{\sqrt{\lambda}}}\right)^2\times\frac{Q^2}{M_{J/\Psi}^2}\right)\times \mathbb N^\prime\,,
\eea
is fixed by our preceding arguments. Strict bulk-to-boundary correspondence implies $\mathbb{N}^\prime=1$ and $\mathcal{N}_{R}=1$ in the double limit of large $N_c$ and strong gauge coupling 
$\lambda$. Here we assume proportionality between the bulk and boundary with $\mathbb{N}^\prime$ and $\mathcal{N}_{R}=1$ overall parameters that capture the finite $N_c$ corrections. They will be fixed 
by the best fit to the data below.

\begin{figure}[!htb]
\includegraphics[height=5.5cm]{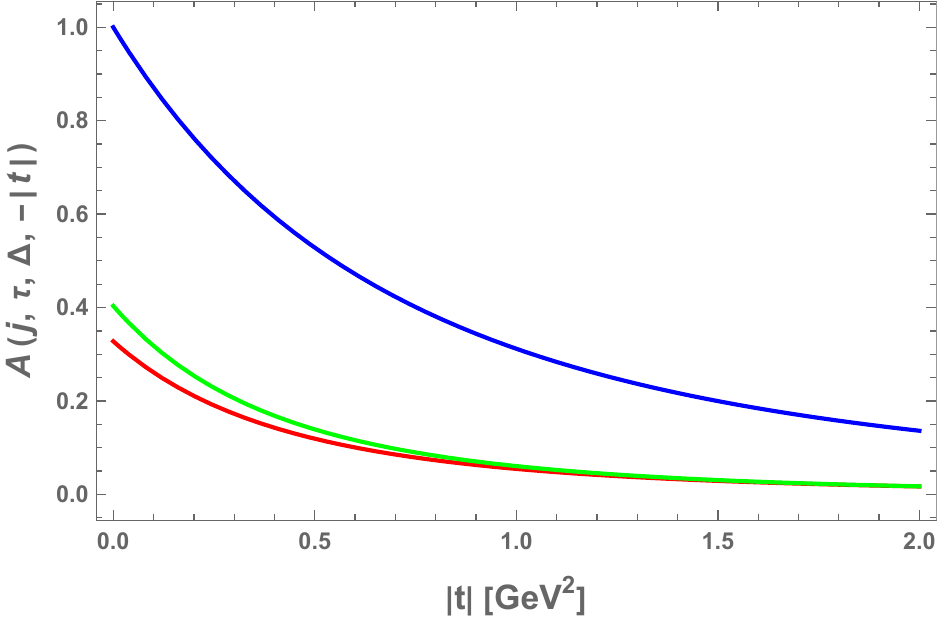}
  \caption{The spin-j nucleon form factor ${\cal A}(j,\tau,\Delta(j),t=-|t|)$: Upper-blue-curve is $j=j_0=2$ and $\Delta(j=2)=4$ in (\ref{FFj22her}); Middle-green curve is $j=j_0=2-2/\sqrt{\lambda}$ 
  and $\Delta(j=j_0)=2$ with $\sqrt{\lambda}=\infty$ in (\ref{FFj22her});
  Lower-red-curve is $j=j_0=2-2/\sqrt{\lambda}$ and $\Delta(j=j_0)=2$ with $\sqrt\lambda=\sqrt{11.243}$ in (\ref{FFj22her}).}
  \label{fig_AJT}
\end{figure}

$\mathcal{A}(j_0,\tau,\Delta,t)$ is the Pomeron-nucleon form factor  ${\mathbb P}+p\rightarrow p$

\bea \label{FFj22her}
&&\mathcal{A}(j=j_0,\tau,\Delta,t)=\Gamma\left(a_K+\frac{\Delta(j)}{2}\right)\nonumber\\
&\times&\frac{2^{1-\Delta }}{\Gamma (\tau )}\Bigg((\tau -1) \Gamma \left(\frac{j}{2}+\tau -\frac{\Delta }{2}\right) \Gamma \left(\frac{1}{2} (j+\Delta +2 \tau -4)\right) \, _2F_1\left(\frac{1}{2} (j-\Delta +2 \tau ),\frac{1}{2} \left(-\Delta +2a_k+4\right);\frac{1}{2} \left(j+2 \tau +2a_k\right);-1\right)\nonumber\\
&+&\Gamma \left(\frac{j}{2}+\tau -\frac{\Delta }{2}+1\right) \Gamma \left(\frac{1}{2} (j+\Delta +2 \tau -2)\right) \, _2F_1\left(\frac{1}{2} (j-\Delta +2 \tau +2),\frac{1}{2} \left(-\Delta +2a_k+4\right);\frac{1}{2} \left(j+2 \tau +2a_k+2\right);-1\right)\Bigg)\,,\nonumber\\
\eea
with the proton twist $\tau=3$ (fixed by the hard counting rule of the proton electromagnetic form factor), $j_0=2-\frac{2}{\sqrt{\lambda}}$, $\Delta(j=j_0)=2$, and $a_K={K^2}/{8\tilde\kappa_N^2}={-t}/{8\tilde\kappa_N^2}$ at $\tilde\kappa_N=0.350~GeV$ (fixed by the mass  of the rho meson and proton). (\ref{FFj22her}) controls the  t-dependence in the high energy regime $\sqrt{s}\gg \sqrt{|t|}$. 

In Fig.~\ref{fig_AJT} we show 
the spin-j nucleon form factor ${\cal A}(j,\tau,\Delta(j),t=-|t|)$ versus $t$. The red-curve is the spin-$j_0$ nucleon form factor or Pomeron-nucleon form factor ${\cal A}(j=j_0,\tau=3,\Delta(j=j_0)=2,t=-|t|)$ with $j_0=2-2/\sqrt\lambda$ and $\sqrt\lambda=\sqrt{11.243}$ in (\ref{FFj22her}).The corresponding mass radius squared is $(0.425~\rm fm)^2$. The green-curve is the spin-$j_0$ nucleon form factor or the Pomeron-nucleon form factor ${\cal A}(j=j_0,\tau=3,\Delta(j=j_0)=2,t=-|t|)$ with $j_0=2-2/\sqrt\lambda$ and $\sqrt\lambda=\infty$ in (\ref{FFj22her}). The corresponding  mass radius squared is $(0.482~\rm fm)^2$. The blue-curve is the spin-$2$ nucleon form factor or the gravitational spin-2 form factor ${\cal A}(j=2,\tau=3,\Delta(j=2)=4,t=-|t|)=\tilde{A}(t=-|t|)$ in (\ref{FFj22her}) with  mass radius squared of $(0.575~\rm fm)^2$ ~\cite{Mamo:2019mka,Mamo:2021krl}. 
The resummed spin-j gluonic sourced by the nucleon  is more compact with a heavier tail.

\subsubsection{Total cross section far from threshold}

The total cross section for electro-production of a vector meson follows from the differential cross section  using  the optical theorem

\be\label{totalcs}
\sigma_{tot}(s,Q^2)=\bigg(\frac{16\pi}{1+\rho^2}\left(\frac{d\sigma}{dt}\right)_{{\rm tot}}\bigg)_{t=0}^{\frac 12}
\ee
with   the rho-parameter

\be
\rho=\frac{\text{Re}[{\cal A}^{tot}_{\gamma* p\rightarrow J/\Psi p} (s,t=0)]}{\text{Im}[{\cal A}^{tot}_{\gamma*  p\rightarrow J/\Psi p} (s,t=0)]}\simeq\frac{\sqrt{\lambda}}\pi\,.
\ee
The cross sections for transversely and longitudinally polarized processes are

\bea
\label{sigmaT}
\sigma_{T}(s,Q^2)=&&\bigg(\frac{16\pi}{1+\rho^2}\frac{d\sigma(s,t,Q,M_{J/\Psi},\epsilon_{T},\epsilon'_{T})}{dt}\bigg)_{t=0}^{\frac 12}\,,\nonumber\\
\sigma_{L}(s,Q^2)=&&\bigg(\frac{16\pi}{1+\rho^2}\frac{d\sigma(s,t,Q,M_{J/\Psi},\epsilon_{L},\epsilon'_{L})}{dt}\bigg)_{t=0}^{\frac 12}\,.
\eea
A more  explicit form of the total cross section  (\ref{totalcs}) is 


\be\label{newtotalcs}
\sigma_{tot}(s,Q^2)=\mathcal{N}_{Q^2}(s,\lambda,f_{J/\Psi},M_{J/\Psi};\mathbb N)\times\mathcal{\tilde{I}}(\lambda,Q,\tilde{\kappa}_{J/\Psi})\times\left(1+\mathcal{N}_{R}^2\times\left(\frac{1}{2-\frac{1}{\sqrt{\lambda}}}\right)^2\times\frac{Q^2}{M_{J/\Psi}^2}\right)^{1/2}\,,\nonumber\\ 
\ee
where $\mathcal{\tilde{I}}(\lambda,Q,\tilde{\kappa}_{J/\Psi})$ is given by (\ref{IQ}), and we have defined

\bea
\mathcal{N}_{Q^2}(s,\lambda,f_{J/\Psi},M_{J/\Psi};\mathbb N)&\equiv&
\bigg({\frac{16\pi}{1+\rho^2}}\bigg)^{\frac 12}\times
\bigg({\left(\frac{s}{\tilde{\kappa}_N^2}\right)^{2-\frac{4}{\sqrt{\lambda}}}\times\mathcal{N}^{TT}(j_0,s,t=0,Q,M_{J/\Psi},m_N)\times \mathcal{A}^2(j_0,\tau,\Delta,t=0)}\bigg)^{\frac 12}\nonumber\\
&\times & \frac{1}{2}\frac{g_5f_{J/\Psi}}{M_{J/\Psi}}\times\left(2-\frac{1}{\sqrt{\lambda}}\right)\times\frac{1}{4}\Gamma^2 \left(2-\frac{1}{\sqrt{\lambda}}\right)\times \mathbb N\,,
\eea
Again, we  assume proportionality between the bulk and boundary with $\mathbb N$ and $\mathcal{N}_{R}$ overall parameters that capture the finite $N_c$ corrections. They will be fixed by the best fit to the data below.

\section{Results for near threshold}

For quasi-real electroproduction ($Q^2\ll \frac{9}{\mathcal{\tilde{N}}_{R}}\times M_{J/\Psi}^2$), we can approximate the normalization (\ref{normdctotalth}) as
\bea\label{normdctotalth2}
\mathcal{N}_t(s,Q;\mathbb{\tilde{N}}^\prime)&\equiv& \mathcal{I}^2(Q,M_{J/\Psi})\times\left(\frac{s}{\tilde{\kappa}_N^2}\right)^{2}\times\mathcal{N}^{TT}(s,t,Q,M_{J/\Psi},m_N)\nonumber\\
&\times& \left(1+\mathcal{\tilde{N}}_{R}^2\times\frac{1}{9}\times\frac{Q^2}{M_{J/\Psi}^2}\right)\times \mathbb{\tilde{N}}^\prime\,.
\eea
In Fig.~\ref{dsigmadtQsquaretth} we show our prediction for the variation of the total differential cross section with $Q^2$ (\ref{newdctotalth}) and fixed $\sqrt s=\sqrt 21~\text{GeV}$, 
in the quasi-real electro-production  regime near threshold ($Q^2\ll \frac{9}{\mathcal{\tilde{N}}_{R}}\times M_{V}^2$), using the normalization (\ref{normdctotalth2}). We have fixed the
$J/\Psi$ parameters as follows: 
$M_{J/\Psi}=3.1~\text{GeV}$, $f_{J/\Psi}=0.405~\text{GeV}$, $\tilde{\kappa}_{J/\Psi}=1.03784~\text{GeV}$ (using the high energy electro-production data for $J/\Psi$ as discussed in the next section), 
and $A^2(0)\times\mathbb{\tilde{N}}^\prime=30.7944~[\text{nb}/\text{GeV}^2]$. The upper-blue-curve is for $Q=0$. The middle-red-curve is for $Q=1.2~\text{GeV}$. The lower-green-curve is for $Q=2.2~\text{GeV}$. 
The dashed-purple-lines  are the total differential cross sections using our kinematic factors using the lattice dipole gravitational form factor with $m_A=1.13~\rm GeV$~\cite{MIT}. 
The data for $Q=0$ is from the GlueX collaboration~\cite{GLUEX}. With increasing $Q$, the differential cross section becomes less sensitive to $|t|$ in the treshold region.

In Fig.~\ref{sigmaWth}  we show the total cross section at the photon-point with $Q^2=0$ for $J/\Psi$ photo-production, by fixing $A^2(0)\times\mathbb{\tilde{N}}=240\pm 47$
and $\tilde{\kappa}_{J/\Psi}=1.03784$ (using the high energy electro-production data for $J/\Psi$ as discussed in the next section). The $J/\Psi$ parameters are fixed as before,
with  $e=0.3$, $\tilde{\kappa}_{N}=0.350~\text{GeV}$ and $m_N=0.94~\text{GeV}$. The data are from GlueX~\cite{GLUEX}.

\begin{figure}[!htb]
\includegraphics[height=5.5cm]{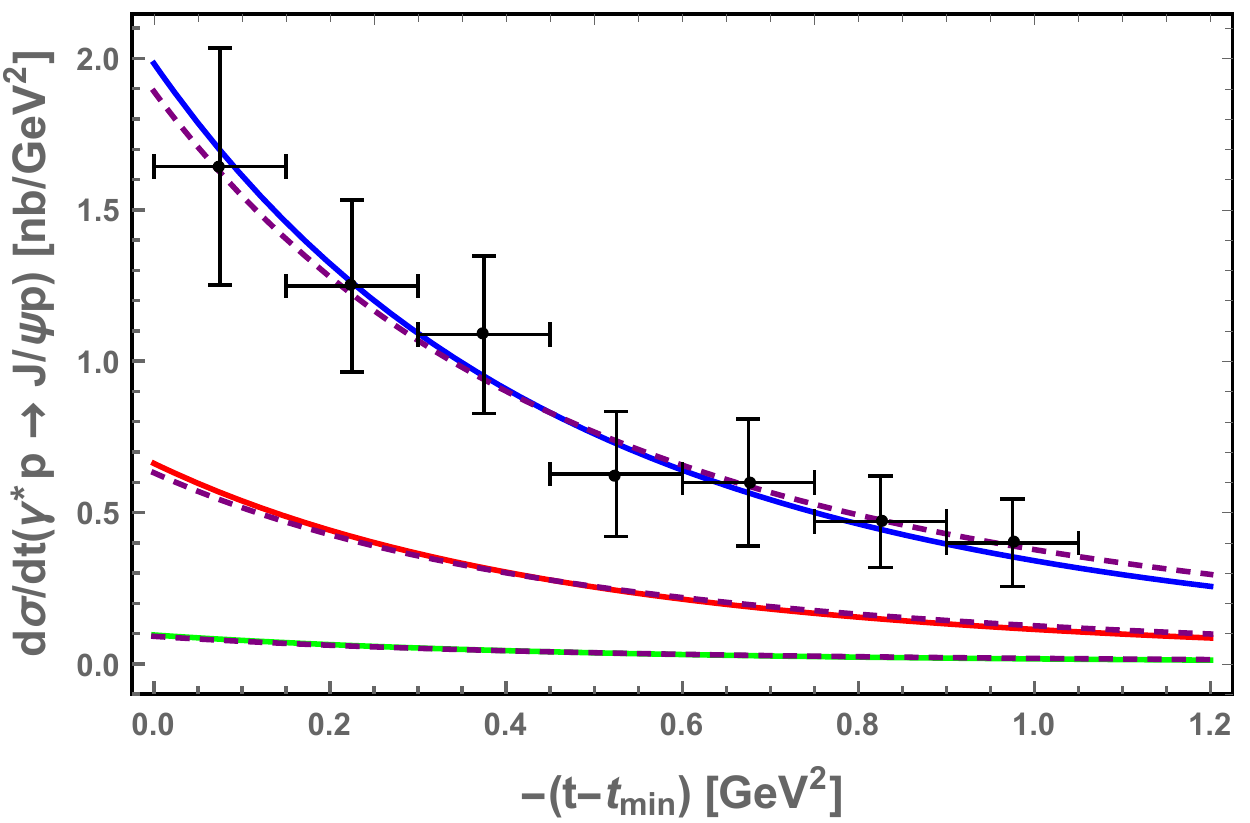}
  \caption{The total differential cross section (\ref{newdctotalth}) normalized using (\ref{normdctotalth2}) for  $\sqrt s=\sqrt 21~\text{GeV}$. The upper-blue-curve is for $Q=0$, 
  the middle-red-curve is for $Q=1.2~\text{GeV}$, and the lower-green-curve is for $Q=2.2~\text{GeV}$. The purple-dashed-curves are our results using 
  the lattice dipole gravitational form factor with $m_A=1.13~\rm GeV$~\cite{MIT}. The data is from the GlueX collaboration at JLab~\cite{GLUEX}.}
  \label{dsigmadtQsquaretth}
\end{figure}

\begin{figure}[!htb]
\includegraphics[height=5.5cm]{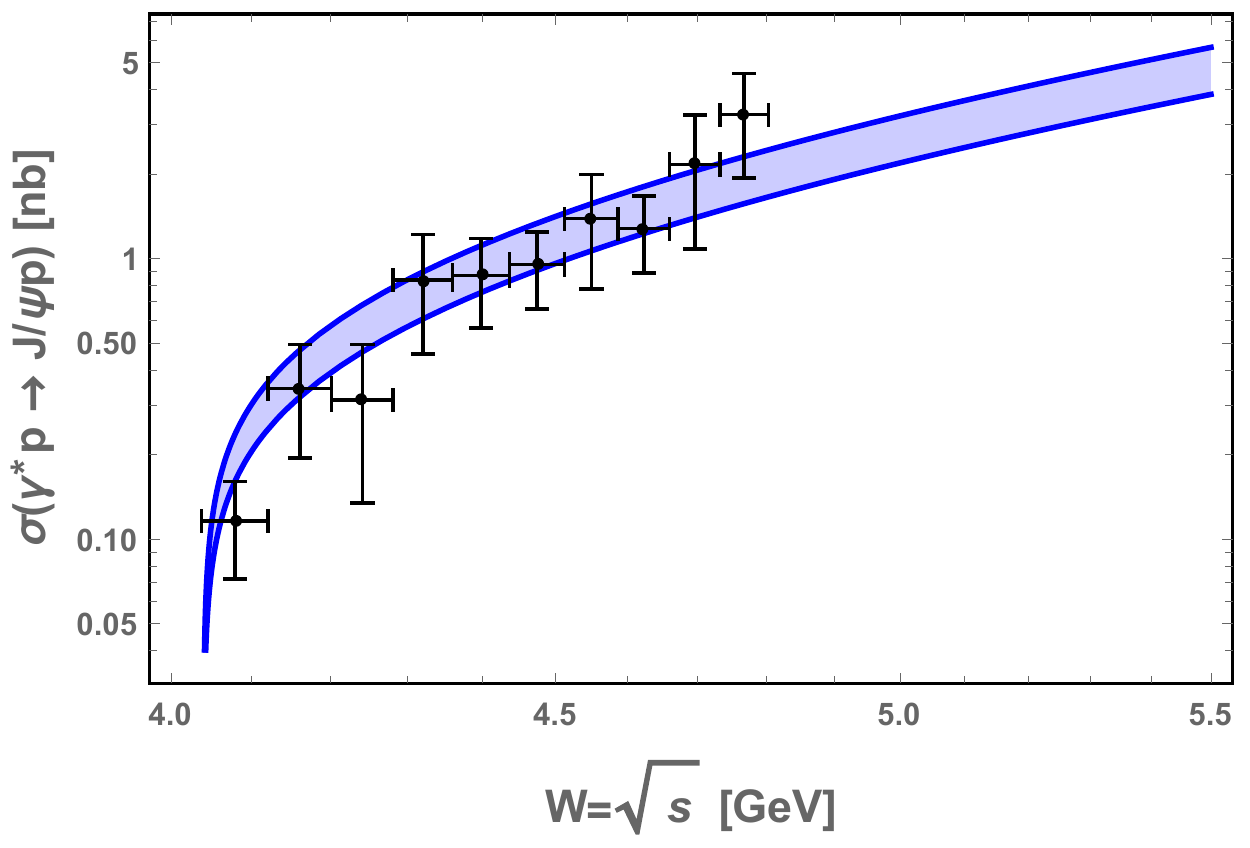}
  \caption{Total holographic cross section  (\ref{newtotalcsth}) for photo-production of $J/\Psi$ versus the GlueX data~\cite{GLUEX}. See text.}
  \label{sigmaWth}
\end{figure}

\begin{figure}[!htb]
\includegraphics[height=5.5cm]{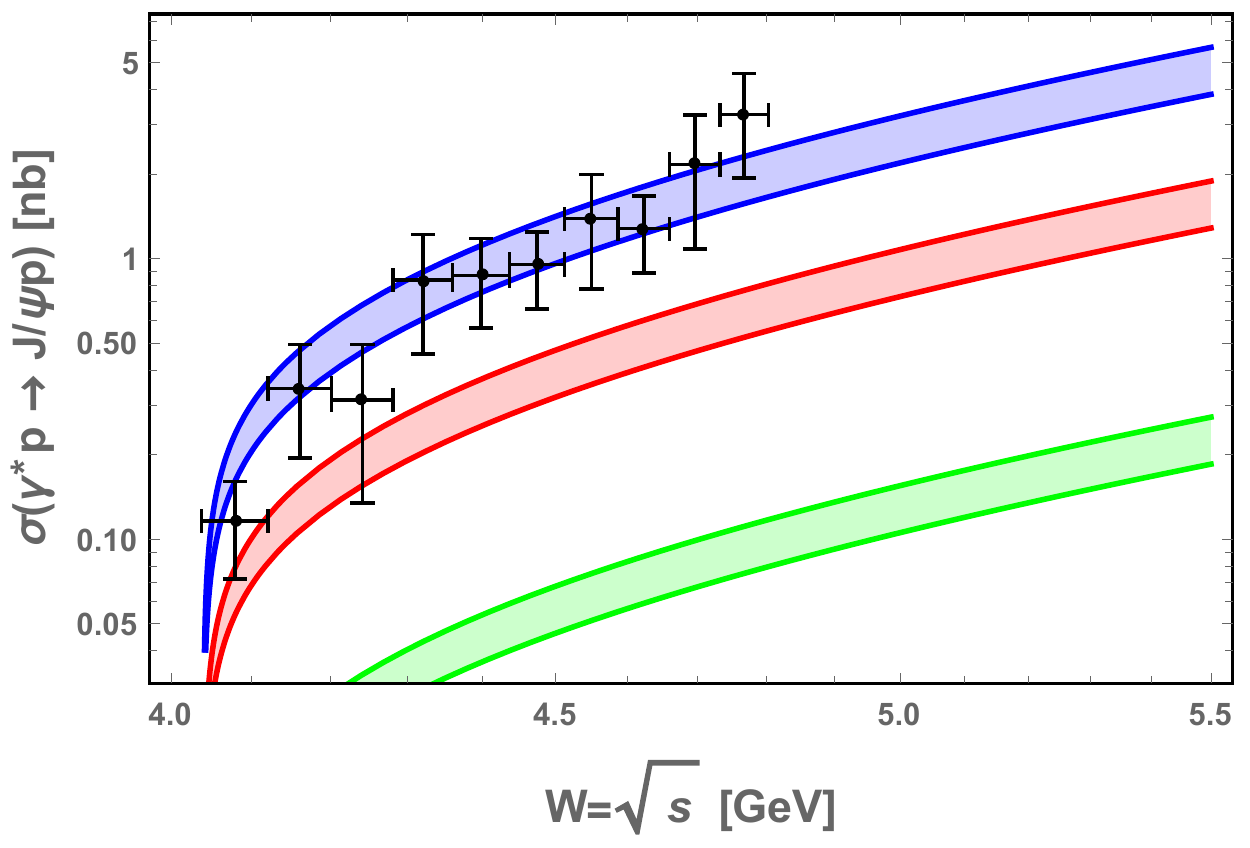}
  \caption{The total cross section (\ref{newtotalcsth2}) 
 for $\tilde{\kappa}_{N}=0.350~\text{GeV}, e=0.3, f_{J/\Psi}=0.405~\text{GeV},M_{J/\Psi}=3.1~\text{GeV}, A^2(0)\times\mathbb{\tilde{N}}=240\pm 47, \tilde{\kappa}_{J/\Psi}=1.03784~\text{GeV}$. The blue band is for $Q^2=0$ (the data is from GlueX (\cite{GLUEX})), the red band is for $Q^2=1.2^2~\text{GeV}^2$, the green band is for $Q^2=2.2^2~\text{GeV}^2$.}
  \label{sigmaWQsquareth}
\end{figure}

For quasi-real electro-production ($Q^2\ll \frac{9}{\mathcal{\tilde{N}}_{R}}\times M_{J/\Psi}^2$), we can approximate the total cross section (\ref{newtotalcsth}) as

\be\label{newtotalcsth2}
\sigma_{tot}(s,Q^2)\approx\mathcal{\tilde{N}}_{Q^2}(s,f_{J/\Psi},M_{J/\Psi};n)\times\mathcal{\tilde{I}}^2(Q,\tilde{\kappa}_{J/\Psi})\,,
\ee
In Fig.~\ref{sigmaWQsquareth} we show~(\ref{newtotalcsth2}) versus $\sqrt{s}$ for $Q^2=0$ in upper-blue-filled-band, 
for $Q^2=1.2^2~\text{GeV}^2$ middle-red-filled-band,  and $Q^2=2.2^2~\text{GeV}^2$ lower-green-filled-band. The data is from GlueX~\cite{GLUEX}.
In Fig.~\ref{sigmaQsquareth} we show the total cross section (\ref{newtotalcsth2}) versus $Q^2$ for fixed $W=\sqrt s=4.4~\text{GeV}$.
Again, we have fixed the $J/\Psi$ parameters as: $e=0.3, f_{J/\Psi}=0.405~\text{GeV}$, $M_{J/\Psi}=3.1~\text{GeV}$  and  $\tilde{\kappa}_{J/\Psi}=1.03784~\text{GeV}$. The bands follow
from the range in  the choice of the overall normalization $A^2(0)\times\tilde{n}=240\pm 47$.


\begin{figure}[!htb]
\includegraphics[height=5.5cm]{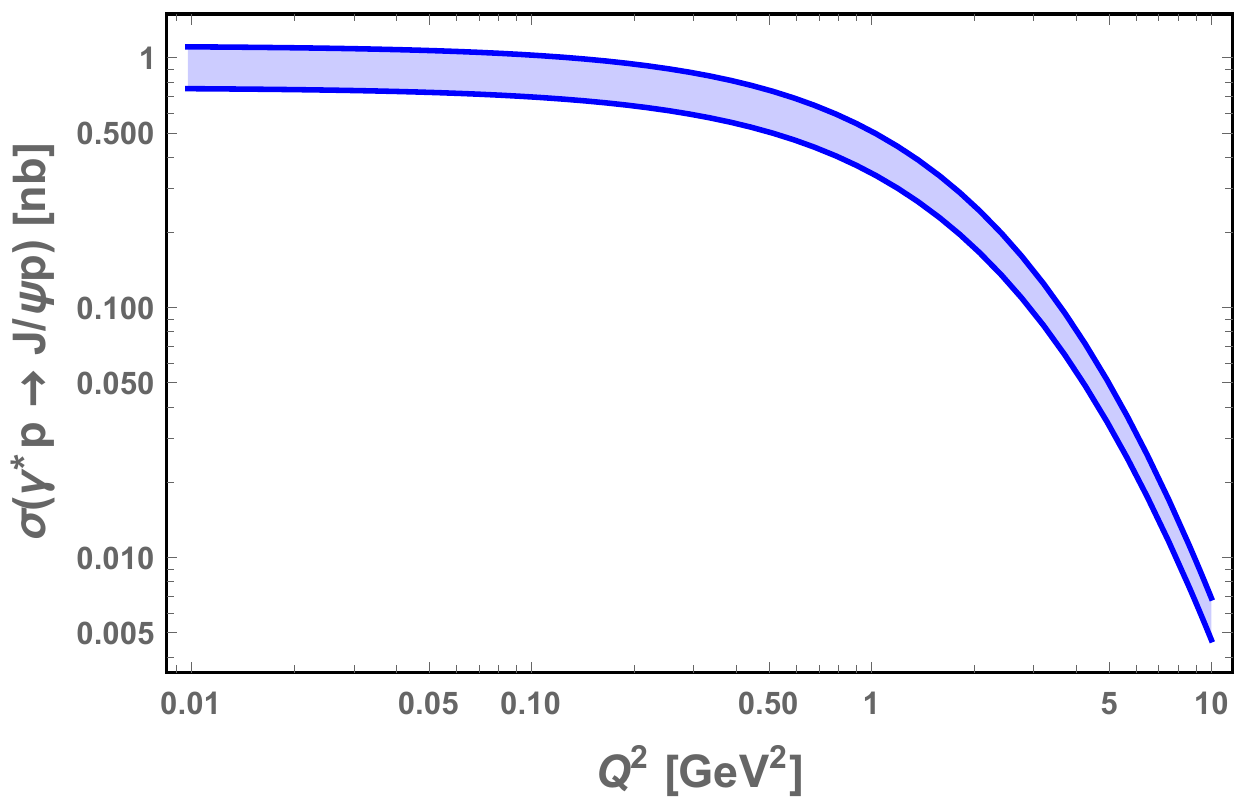}
  \caption{The total cross section for quasi-real electro-production of $J/\Psi$ in (\ref{newtotalcsth2}) versus $Q^2$ for fixed $W=\sqrt s=4.4~\text{GeV}$. See text.}
  \label{sigmaQsquareth}
\end{figure}

\section{Results  far from threshold}

\subsection{$J/\Psi$ electro-production}

In the  $J/\Psi$ electro-production channel, we fix the $^\prime$t Hooft coupling $\lambda=11.243$ and $\tilde\kappa_{J/\Psi}=1.03784~\text{GeV}$,
for a mass $M_{J/\Psi}=3.1$ GeV and a decay constant  $f_{J/\Psi}=0.405$ GeV.  In Fig.~\ref{RLT-PSI} we show the $Q^2$ dependence of the ratio of the longitudinal 
 to transverse  cross sections as in (\ref{sigmaT})

\be
\label{RE1}
R=\frac{\sigma_L}{\sigma_R}=\mathcal{N}_R\times \left(\frac{1}{2-\frac{1}{\sqrt{11.243}}}\right)\times\frac{Q}{3.1~\text{GeV}}
\ee
The overall and arbitrary normalization is ${\cal N}_R=0.6$ for the blue-solid curve.
The data   are  from the 2005 H1 collaboration at HERA~\cite{Aktas:2005xu}. The slow rise in the semi-logarithmic plot is  consistent 
with the linear Q-behavior following from holography, since $d\sigma_L/dt\sim 1/Q^6$ and  $d\sigma_T/dt\sim 1/Q^8$. Asymptotically,
the effective size is fixed by the virtual photon size $1/Q$. 

\begin{figure}[!htb]
\includegraphics[height=5.5cm]{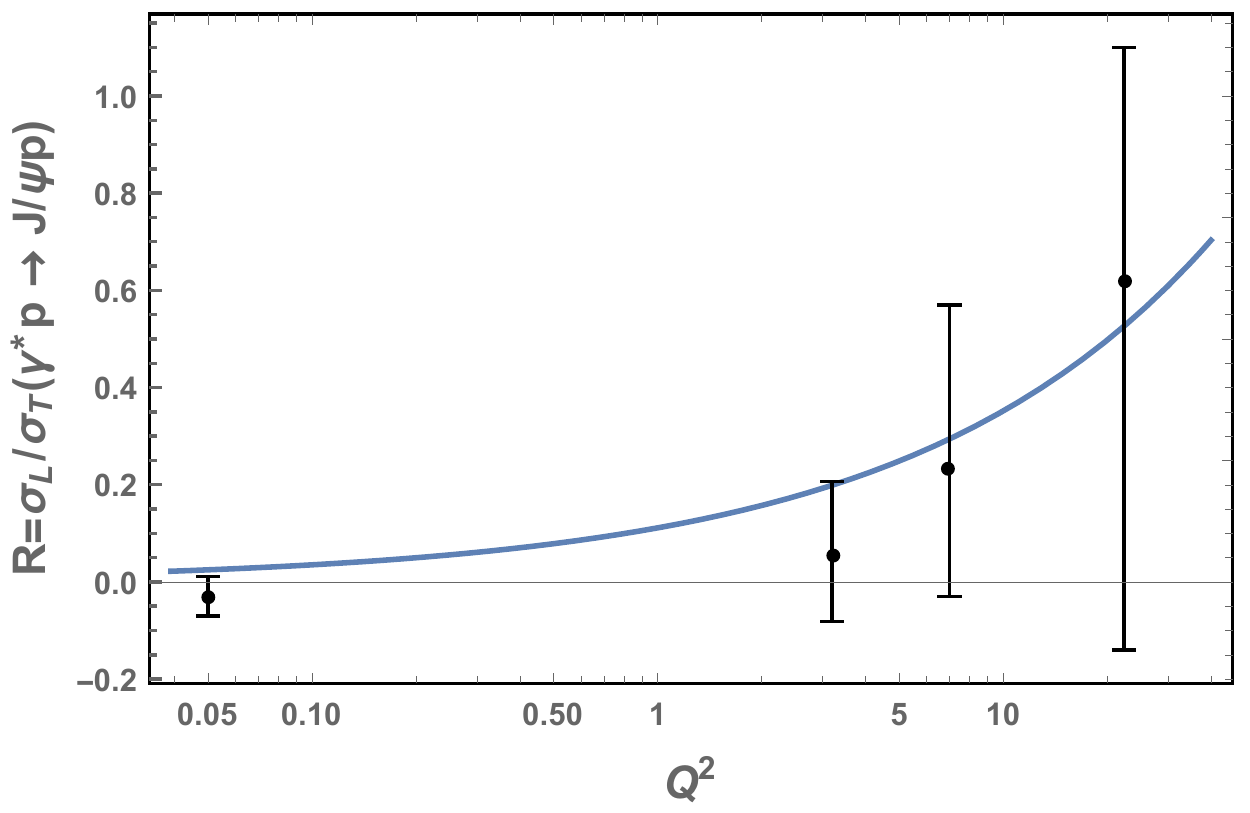}
  \caption{Ratio $R=\sigma_L/\sigma_T$ of the total longitudinal to transverse cross sections versus $Q^2$  for $J/\Psi$ electro-production as given in (\ref{RE1}).
  }
  \label{RLT-PSI}
\end{figure}

\begin{figure*}
\subfloat[\label{xuatqq6pt5goined300pdfsetssmallx}]{%
  \includegraphics[height=6cm,width=.49\linewidth]{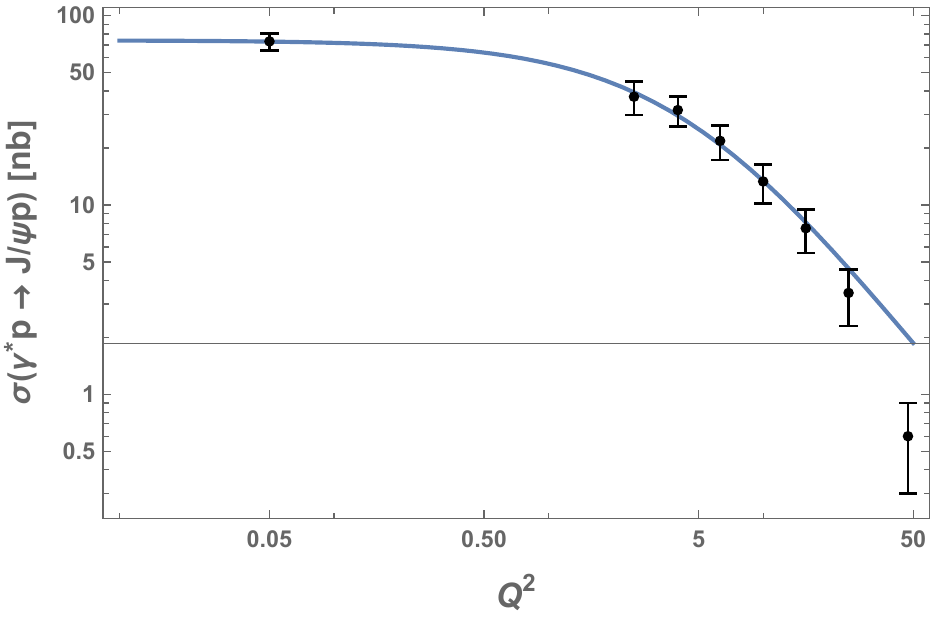}%
}\hfill
\subfloat[\label{xdatqq6pt5goined300pdfsetssmallx}]{%
  \includegraphics[height=6cm,width=.49\linewidth]{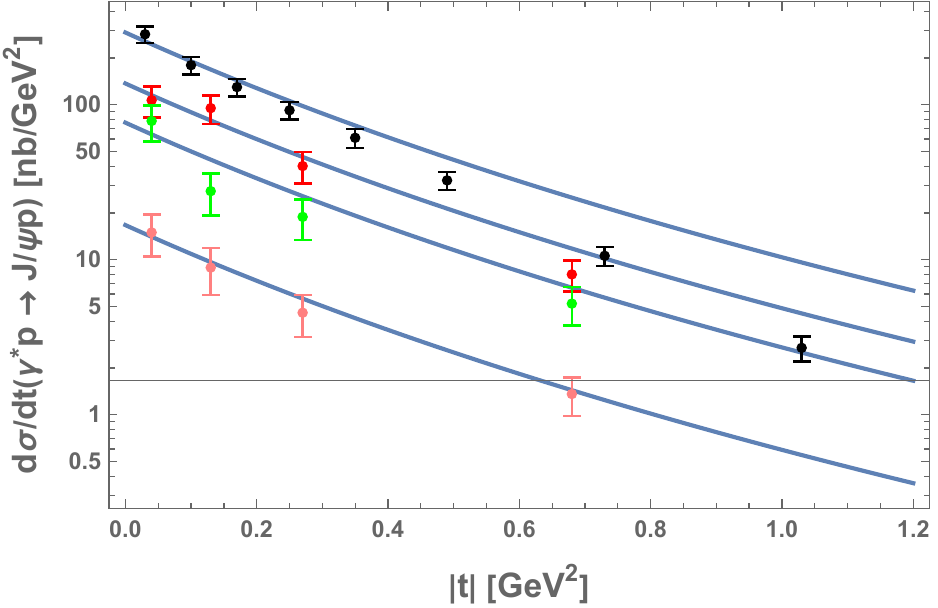}%
}
\caption{a: the blue-solid line is the total $J/\Psi$ electro-production cross section versus $Q^2$ for fixed $\sqrt{s}=90$ GeV, in comparison to the HERA data~\cite{Aktas:2005xu};
(b): the blue-solid lines are the differential $J/\Psi$ electro-production cross section versus $|t|$,  for $40\,{\rm GeV}<\sqrt{s}<160\,{\rm GeV}$ 
in comparison to the HERA data~\cite{Aktas:2005xu}.  The black-data points are for $Q=\sqrt{0.05}\,{\rm GeV}$, the red-data points are for $Q=\sqrt{3.2}\,{\rm GeV}$, the green-data points are for $Q=\sqrt{7}\,{\rm GeV}$, the pink-data points are for $Q=\sqrt{22.4}\,{\rm GeV}$. }
\label{TOTAL-JPSII}
\end{figure*}

In Fig.~\ref{TOTAL-JPSII}a, we show the total holographic cross section for the electro-production of $J/\Psi$ as given  in  (\ref{newtotalcs}) versus $Q^2$
and for fixed $\sqrt{s}=90$ GeV,
\be
\sigma_{tot}(Q^2)=\mathcal{N}_{Q^2}\times\mathcal{\tilde{I}}(\lambda=11.243,Q,\tilde{\kappa}_{J/\Psi}=1.03784~\text{GeV})\times\left(1+0.6^2\times\left(\frac{1}{2-\frac{1}{\sqrt{11.243}}}\right)^2\times\frac{Q^2}{3.1^2~\text{GeV}^2}\right)^{1/2}
\ee
with the form factor $\mathcal{\tilde{I}}(\lambda,Q,\tilde{\kappa}_{J/\Psi})$ given in (\ref{IQ}), $\lambda=11.243, f_{J/\Psi}=0.405~\text{GeV}, M_{J/\Psi}=3.1~\text{GeV}, A^2(0)\times {\mathbb N}=206,556~[nb]$, $\mathcal{N}_R=0.6$, and $\tilde{\kappa}_{J/\Psi}=1.03784~\text{GeV}$.
The holographic cross section is in agreement with the reported data in the range  $0\leq Q^2\leq 10\,{\rm GeV}^2$.
This is expected since at higher $Q^2$ the weak coupling regime sets in. This observation is consistent with our recent
analysis of neutrino-nucleon DIS scattering in holographic QCD~\cite{Mamo:2021cle}. Fig.~\ref{TOTAL-JPSII}a shows that the
holographic $Q^2$ dependence following from $\tilde{\cal{I}}$ is consistent with the data at low and intermediate values of $Q^2$.
This dependence originates  from the bulk-to-boundary vector propagator (\ref{IJJ}) which re-sums the $c\bar c$ radial Regge trajectory.

In Fig.~\ref{TOTAL-JPSII}b we show the differential cross section  (\ref{newdctotal}) for  the electro-production of $J/\Psi$,
after adjusting the overall matching parameter $\mathbb N^\prime$ to a data point, as follows

\bea
&&40^2~{\rm GeV}^2<s<160^2~{\rm GeV}^2, Q=\sqrt{0.05}~{\rm GeV}, A^2(0)\times{\mathbb N}^{\prime}=9.33\times 10^{13}~[nb/GeV^2]\qquad {\rm black-data}, \nonumber\\
&&40^2~{\rm GeV}^2<s<160^2~{\rm GeV}^2, Q=\sqrt{3.2}~{\rm GeV}, A^2(0)\times{\mathbb N}^{\prime}=1.98\times 10^{14}~[nb/GeV^2]\qquad\,\,\, {\rm red-data}, \nonumber\\
&&40^2~{\rm GeV}^2<s<160^2~{\rm GeV}^2,Q=\sqrt{7}~{\rm GeV}, A^2(0)\times{\mathbb N}^{\prime}=3.62\times 10^{14}~[nb/GeV^2] \qquad \,\,\,\,\,\,\,{\rm green-data}, \nonumber\\
&&40^2~{\rm GeV}^2<s<160^2~{\rm GeV}^2,Q=\sqrt{22.4}~{\rm GeV}, A^2(0)\times{\mathbb N}^{\prime}=1.01\times 10^{15}~[nb/GeV^2]\qquad  {\rm pink-data}.\nonumber\\
\eea
The data are from the 2005 H1 collaboration at HERA in~\cite{Aktas:2005xu}. The $|t|$ dependence of the holographic differential cross sections  follow from
$|{\cal A}(j_0, \tau, \Delta, t)|^2$ as given in  (\ref{FFj22her}), and is consistent with the reported data for $J/\Psi$ electro-production at HERA for different $Q^2$. Recall that 
in the threshold region with $\sqrt{s}\sim (m_N+m_{J/\Psi})$,  ${\cal A}(2,3,4,t)=A(t)$  in (\ref{FFj22}) which is the gravitational tensor coupling.  Its Reggeized form 
${\cal A}(j_0,\tau,\Delta,t)$ is probed by the differential cross section of $J/\Psi$ electro-production at large $\sqrt{s}$.

\subsection{$\phi$ electro-production}

Although most of our arguments and derivations are justified  for the electro-production of the heavy meson $J/\Psi$ (and even better for $\Upsilon$), we can minimally
adjust them to describe the lighter $\phi$ meson, with a weaker justification. 
The kinematically adjusted ratio  of the longitudinal to transverse (\ref{sigmaT}) total cross sections 
for $\phi$ electro-production (after replacing $M_{J/\Psi}$ by $M_{\phi}$) is

\be
\label{RE2}
R=\frac{\sigma_L}{\sigma_R}=\mathcal{N}_R\times \left(\frac{1}{2-\frac{1}{\sqrt{11.243}}}\right)\times\frac{Q}{1.019~\text{GeV}}
\ee
where the arbitrary normalization coefficient $\mathcal{N}_R=2.6$ for the blue-solid-curve
in Fig.~\ref{SIGMA-PHI-RHO}.
 The data  are  from the 2010 H1 collaboration in~\cite{Aaron:2009xp}. Again, on the semi-logarithmic scale, the rise with $Q^2$ 
 supports the linear ratio (\ref{RE2}) in the expected range of validity of $Q^2$, although more data in this range would be
 welcome.

\begin{figure*}
\subfloat[\label{xdatqq6pt5goined300pdfsetssmallx}]{%
  \includegraphics[height=6cm,width=.49\linewidth]{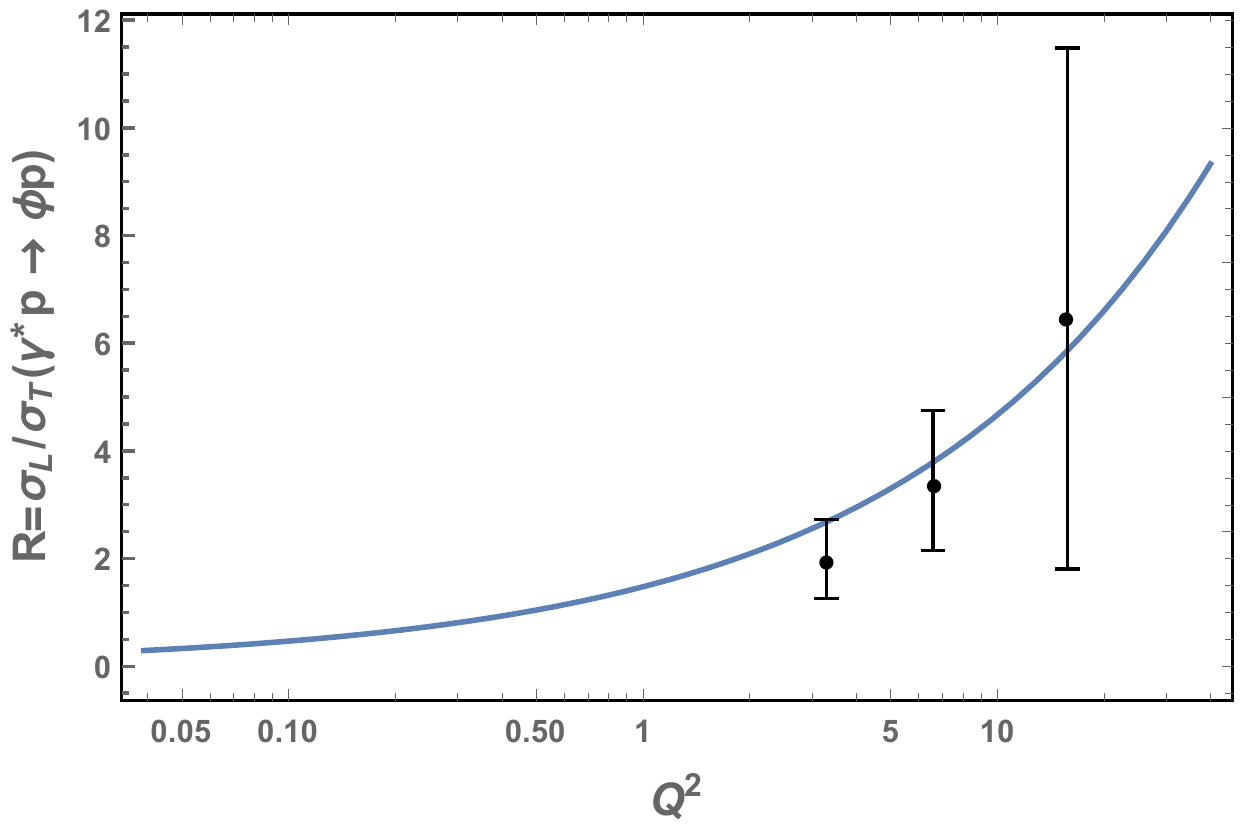}%
}\hfill
\caption{Ratio $R=\sigma_L/\sigma_T$ of the total longitudinal to transverse cross sections versus $Q^2$,  for $\phi$ electro-production as given in (\ref{RE2}). The blue-solid-curve is for ${\cal N}_R=2.6$. 
The data are from the H1 collaboration~\cite{Aaron:2009xp}}
\label{SIGMA-PHI-RHO}
\end{figure*}

\begin{figure*}
\subfloat[\label{xuatqq6pt5goined300pdfsetssmallx}]{%
  \includegraphics[height=6cm,width=.49\linewidth]{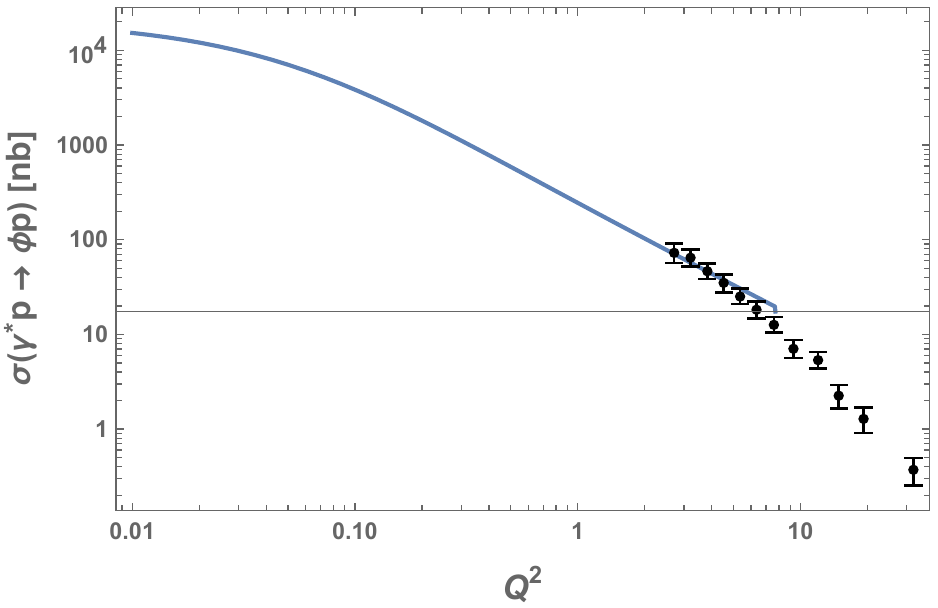}%
}\hfill
\subfloat[\label{xdatqq6pt5goined300pdfsetslargex}]{%
  \includegraphics[height=6cm,width=.49\linewidth]{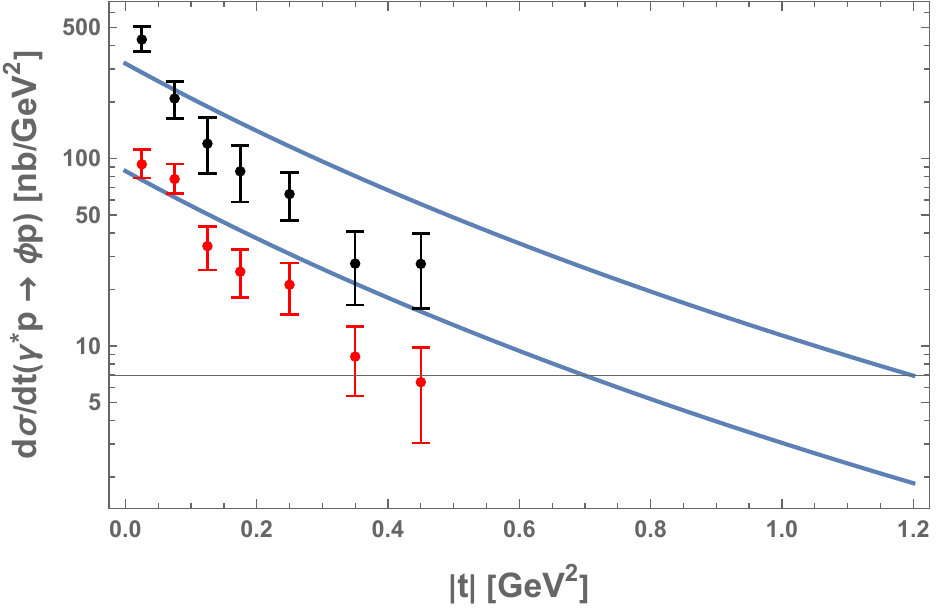}%
}\hfill
\caption{a: total cross section for $\phi$ electro-production versus $Q^2$ for $\sqrt{s}=75\,{\rm GeV}$; 
b: differential cross section  for $\phi$ electro-production versus $|t|$
 for $\sqrt{s}=75\,{\rm GeV}$ and $Q^2=3.3\,{\rm GeV}^2$ (black-data) and $Q^2=6.6\,{\rm GeV}^2$ (red-data).
}
\label{PHIRHO}
\end{figure*}

The adjusted total cross section for $\phi$ electro-production follows from (\ref{newtotalcs})

\be
\label{RE4X}
\sigma_{tot}(Q^2)=\mathcal{N}_{Q^2}\times\mathcal{\tilde{I}}(\lambda=11.243,Q,\tilde{\kappa}_{\phi}=0.1042~\text{GeV})\times\left(1+2.6^2\times\left(\frac{1}{2-\frac{1}{\sqrt{11.243}}}\right)^2\times\frac{Q^2}{1.019^2~\text{GeV}^2}\right)^{1/2}\nonumber\\
\ee
where $\mathcal{\tilde{I}}(\lambda,Q,\tilde{\kappa}_{\phi})$ is given in (\ref{IQ}) (after replacing $J/\Psi$ by $\phi$), $\mathcal{N}_{R}=2.6$,

\be
s=75^2~\text{GeV}^2,\lambda=11.243,f_{\phi}=0.233~\text{GeV}, M_{\phi}=1.019~\text{GeV}, A^2(0)\times\mathbb N=3,624.56~[nb]\,,
\ee
and $\tilde{\kappa}_{\phi}=0.1042~\text{GeV}$ adjusted to the $\phi$-mass.

In Fig.~\ref{PHIRHO}a we show (\ref{RE4X}) versus $Q^2$ for $\sqrt{s}=75\,{\rm GeV}$.
The data are from the 2010 H1 collaboration at HERA in~\cite{Aaron:2009xp}.  In the range $Q^2< 8\,{\rm GeV}^2$, the dependence in
$Q^2$ supports that from the bulk-to-boundary propagator (\ref{IJJ}) (with the pertinent substitutions) over a decade.
The deviation for $Q^2\geq 8\,{\rm GeV}^2$ signals the on-set of weak coupling as systematically noted in our recent neutrino
DIS analysis~\cite{Mamo:2021cle}.
In Fig.~\ref{PHIRHO}b we show the differential cross section for $\phi$ electro-production versus $|t|$, following from
 (\ref{newdctotal}), after replacing $J/\Psi$ by $\phi$,  with

\be
&&s=75^2~{\rm GeV}^2,Q=\sqrt{3.3}~{\rm GeV}, A^2(0)\times {\mathbb N}^{\prime}=1.98\times 10^{10}~[nb/GeV^2]\qquad {\rm black-data}\nonumber\\
&&s=75^2~{\rm GeV}^2,Q=\sqrt{6.6}~{\rm GeV}, A^2(0)\times {\mathbb N}^{\prime}=2.89\times 10^{10}~[nb/GeV^2]\qquad {\rm red-data}
\ee
The data are from the 2010 H1 collaboration at HERA in~\cite{Aaron:2009xp}. The deviations are substantial for moderate values of $t$, which is an indication that Reggeon-like couplings are important in the electro-production of lighter mesons such as $\phi$.

\end{widetext}

\section{Conclusions}

The holographic photo-production of charmonium has shown that at treshold the differential cross section probes the
tensor gravitational coupling. Away from treshold, the tensor Reggeizes to a Pomeron and the differential cross section
probes the Reggeized coupling. We have now extended this analysis to the electro-production of charmonium with a
similar observation for the differential cross section.

In the double limit of large $N_c$ and strong coupling, only
the tensor coupling or A-form  factor drives  the production of $J/\Psi$ (and also $\Upsilon$)  in the treshold region, and its Reggeized form
way above  treshold.
A comparison to the available data at low and intermediate
$Q^2$ shows that the holographic t-dependence for the total differential cross section for charmonium photo- and electro-production, is
 consistent with the recent GlueX data and the HERA data over a broad range of $t|$.


In holography, the $Q^2$ dependence of the electro-production of charmonium  follows from the bulk-to-boundary propagator
sourced by the $1^{--}$ current at the boundary such as $\bar c\gamma_\mu c$.  As a result, the longitudinal differential cross section 
asymptotes $1/Q^6$, and the transverse differential cross section asymptotes $1/Q^8$, in agreement with the 
hard scattering rules.  At smaller $Q^2$, the transverse differential cross section is about constant, while the 
longitudinal differential cross section vanishes as $Q^2$.  This limit is fixed by  the finite size of charmonium.
The holographic results compare well with the HERA data for low and intermediate $Q^2$, but depart from the data at
larger $Q^2$ with the on-set of weak coupling and scaling. 

Our results extend miniminally to the $\phi$-channel. The total cross section and the
ratio of the longitudinal to transverse cross sections are reasonably well reproduced at low and intermediate 
$Q^2$, but the $|t|$ dependence of the differential cross section is slightly off. This is an indication that Reggeon 
exchange is likely important in this channe. More data with better accuracy would be welcome.

\section{Acknowledgements}

K.M. is supported by the U.S.~Department of Energy, Office of Science, Office of Nuclear Physics, contract no.~DE-AC02-06CH11357, and an LDRD initiative at Argonne National Laboratory under Project~No.~2020-0020. I.Z. is supported by the Office of Science, U.S. Department of Energy under Contract No. DE-FG-88ER40388.

\section{\label{kinematics} kinematics of  the $\gamma^*p\rightarrow Vp$ process}

We start by briefly reviewing  the kinematics for the
process $\gamma^*p\rightarrow Vp$.
We first define the Lorentz scalars as $s=W^2=(p_1+q
_1)^2$, and $t=(p_1-p_2)^2=(q_1-q_2)^2$ where $q_{1,2}$ are the four-vectors of the virtual photon and vector meson, respectively (note that we occasionally use the notation $q\equiv q_1$ and $q'\equiv q_2$), and $p_{1,2}$ are the four vector of the proton. Throughout
we will work with mostly negative signature, i.e., $\eta_{\mu\nu}=(+1,-1,-1,-1)$. Note that our convention is different from the mostly positive signature used in most
holographic analyses.

\begin{figure}[!htb]
\includegraphics[height=5cm]{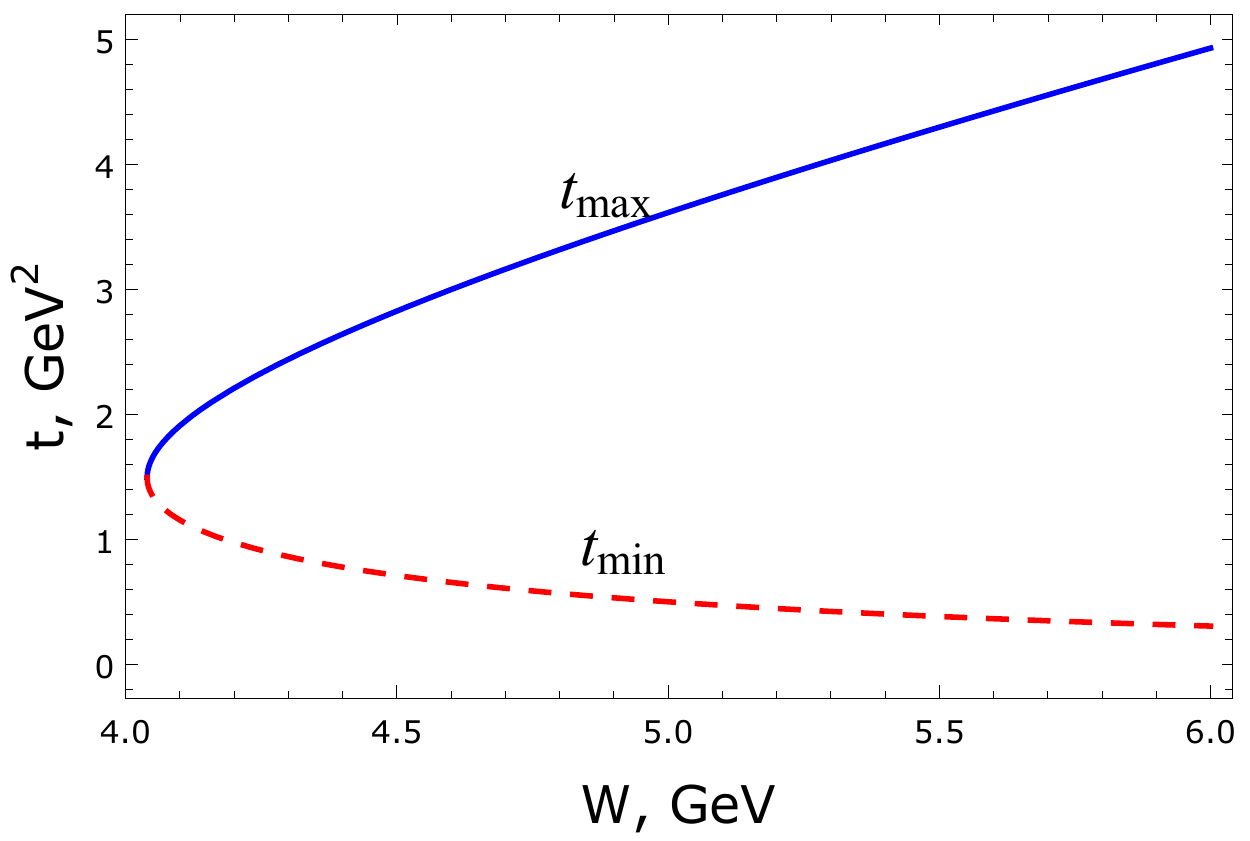}
  \caption{$t_{min}$ and $t_{max}$ vs $W=\sqrt{s}$ for $M_{V}=M_{J/\psi}=3.10~GeV$, $m_{N}=0.94~GeV$, and $Q=0$. Note that at the threshold energy $W_{tr}=\sqrt{s_{tr}}=m_{N}+M_{V}=4.04~GeV$, we have $t_{min}=t_{max}$.}
\label{tmx}
\end{figure}

We will work in the center-of-mass (CM) frame of the pair composed of the virtual photon $\gamma^*$ and the proton. In this frame, one can derive the mathematical relationships between the three-momenta of the virtual photon and vector meson ($\mathbf{q}_{\gamma}$, $\mathbf{q}_V$) and Lorentz scalars ($s$, $t$, $q_1^2=-Q^2$, $q_2^2=M_{V}^2$, $p_1^2=p_2^2=m_{N}^2$) as (see, for example, Eqs.11.2-4 in~\cite{Srednicki:2007qs})

\be
\vert\mathbf{q}_\gamma\vert=\frac{1}{2\sqrt{s}}\sqrt{s^2-2(-Q^2+m_N^2)s+(-Q^2-m_N^2)^2}\,,\nonumber\\
\ee
\be
\vert\mathbf{q}_V\vert=\frac{1}{2\sqrt{s}}\sqrt{s^2-2(M_V^2+m_N^2)s+(M_V^2-m_N^2)^2}\,,\nonumber\\
\ee
and

\be
\label{tminmax}
t=-Q^2+M_V^2-2E_{\gamma}E_V+2\vert\mathbf{q}_\gamma\vert\vert\mathbf{q}_V\vert\cos\theta\,,\nonumber\\
\ee
Here  $E_{\gamma}=(-Q^2+\mathbf{q}_{\gamma}^2)^{\frac 12}$ is the energy of the virtual photon, and
$E_{V}=(M_V^2+\mathbf{q}_{V}^2)^{\frac 12}$ is the energy of the vector meson. The t-transfer at low $\sqrt{s}$
is bounded by $t_{min}\equiv |t\vert_{\cos\theta=+1}|$ and $t_{max}\equiv|t\vert_{\cos\theta=-1}|$
as illustrated in Fig.~\ref{tmx}.

We now note that at threshold and for example $V=J/\Psi$ with $s_{\rm tr}=(m_N+M_V)^2=4.04\,{\rm GeV}^2$

 \bea
 -t_{\rm min}(s=s_{\rm tr})=&&\frac{m_NM^2_V}{m_N+M_V} \nonumber\\
 =&&1.5^2\,{\rm GeV}^2\ll 4.04^2\,{\rm GeV}^2=s_{\rm tr}\nonumber\\
 \label{6}
 \eea
 and away from threshold

 \be
- t_{\rm min}(s\gg s_{\rm tr})\sim \left(\frac {m_NM_V}s\right)^2\ll s
 \label{7}
 \ee
 The electro-production kinematics for charmonium and also  bottomium, is dominated
 by the diffractive process all the way to threshold.




\section{holographic model}

 We consider AdS$_5$  with a background metric  $g_{MN}=(\eta_{\mu\nu},-1)/z^2$ and $\eta_{\mu\nu}=(1,-1,-1,-1)$.
 Confinement will be described by  a background dilaton $\phi=\tilde{\kappa}_{V}^2z^2$ for mesons,
 $\phi=\tilde{\kappa}_{N}^2z^2$ for protons and  $\phi=2\tilde{\kappa}_{N}^2z^2$ for glueballs in the soft wall model.
The glueballs will be described by $h_{\mu\nu}$ and the scalar-dilaton by $\varphi$.
The flavor gauge fields will be described by U(1) gauge fields, and the spin-$\frac 12$ Dirac
 fermion by  $\Psi$.

 \subsection{Bulk Dirac fermion and vector meson}

The bulk Dirac fermion action in curved AdS$_5$ with minimal coupling to the U(1) vector meson is ~\cite{CARLSON}

\be
S=\int d^{5} x \sqrt{g}\,\big(\mathcal{L}_F+\mathcal{L}_V \big)+\int d^4 x \sqrt{-g^{(4)}}\mathcal{L}_{UV}\,,\nonumber\\
\label{Action}
\ee
with the fermionic, gauge field and boundary actions

\bea
\label{fermionAction}
\mathcal{L}_F=&& \frac{1}{2g_{5}^2}e^{-\phi(z)} \nonumber\\
&&\times \bigg( \frac{i}{2} \bar{\Psi} e^N_A \Gamma^A\big(\overrightarrow{D}_N-\overleftarrow{D}_N\big)\Psi-(M+V(z))\bar{\Psi}\Psi\bigg)\,,\nonumber\\
\mathcal{L}_V= &&-\frac{1}{4g^2_5}\,e^{-\phi(z)}\,g^{\mu\alpha}g^{\beta\nu}\,F_{\mu\nu}^V\,F_{\alpha\beta}^V\,,\nonumber\\
\mathcal{L}_{UV}=&&
 \frac{1}{2g_5^2}\left(\bar{\Psi}_L \Psi_R +\bar{\Psi}_R \Psi_L\right)_{z=\varepsilon}\,,
\ee
We have fixed the potential $V(z)=\tilde{\kappa}_{N}^2z^2$ for both the  soft wall model.
We have denoted by  $e^N_A=z \delta^N_A$ the inverse vielbein, and defined the covariant derivatives

\bea
\overrightarrow{D}_N=&&\overrightarrow{\partial}_N +\frac{1}{8}\omega_{NAB}[\Gamma^A,\Gamma^B]-iV_N\nonumber\\
\overleftarrow{D}_N=&&\overleftarrow{\partial}_N +\frac{1}{8}\omega_{NAB}[\Gamma^A,\Gamma^B]+iV_N
\eea
The components of the spin connection are $\omega_{\mu z\nu}=-\omega_{\mu\nu z}=\frac{1}{z}\eta_{\mu\nu}$, the Dirac gamma matrices  satisfy anti-commutation relation $\{\Gamma^A,\Gamma^B\}=2\eta^{AB}$, that is, $\Gamma^A=(\gamma^\mu,-i\gamma^5)$, and $F^V_{MN}=\partial_M V_N-\partial_N V_M$.
The  equation of motions for the bulk Dirac fermion and the U(1) gauge field follow by variation

\bea
&&\big[i e^N_A \Gamma^A D_N -\frac{i}{2}(\partial_N\phi)\, e^N_A \Gamma^A- (M+\phi(z))\big]\Psi =0\,,\nonumber\\
&&\frac{1}{\sqrt{g}}\partial_M\big(\sqrt{g}e^{-\phi}F^{MN}\big)=0\,.\nonumber\\
\eea

The coupling $g_5$ is inherited from the nature of the brane embeddings in bulk:
${1}/{g_{5}^2}\equiv{3N_cN_f}/(12\pi^2)$ (D7-branes),
and ${1}/{g_{5}^2}\equiv ({3\sqrt{\lambda}}/{2^{5/2}\pi}){N_cN_f}/({12\pi^2})$ (D9-branes).
The brane embeddings
with $N_f=1$ are more appropriate for describing heavy mesons in bulk, as the U(1) field mode decompose in
an infinite tower of massive vector  mesons on these branes as we discussed above.
When ignoring these embedding, the standard assignement is:
${1}/{g_{5}^2}\equiv{N_c}/(12\pi^2)$.

We note that in (\ref{fermionAction}), we  have excluded a Yukawa-type coupling between the scalar-dilaton  $\varphi$ and  the bulk Dirac fermion $\Psi$,  since neither the fermionic part of the Type IIB supergravity action (see, for example, Eq.~A.20 in \cite{DHoker:2016ncv}) nor the fermionic part of the  DBI action in string theory (see, for example, Eq.~56 in \cite{Kirsch:2006he}) support such a coupling.

\subsection{Spectra}

For the soft wall model, the vector meson spectrum follows from the equation of motion for $V^N$.
 The results for the heavy meson masses and decay constants are~ \cite{Grigoryan:2010pj}

\bea
\label{SOFTMF}
m_n^2=&&4\tilde\kappa^2_V(n^*+1)\nonumber\\
g_5f_n=&&\sqrt{2}\tilde\kappa_V\left(\frac{n+1}{n^*+1}\right)^{\frac 12}
\eea
with $n^*=n+c_V^2/4\tilde\kappa_V^2$. The additional constant $c_V$ is
fixed as ${c_{V}^2}/{4\tilde{\kappa}_{V}^2}={M_{V}^2}/{4\tilde{\kappa}_{V}^2}-1$ for $n=0$ for the heavy mesons  $V=(J/\psi , \Upsilon)$,
and $c_\rho=0$ for the light mesons. The mass spectrum of the bulk Dirac fermions is given by~\cite{CARLSON}

\be
\label{SOFTMFD}
m_n^2=&&4\tilde\kappa^2_N(n+\tau-1)\,,
\ee
with the  twist factor $\tau=3$.  For the specific soft wall applications
to follow we will set $\tilde\kappa_N=\tilde\kappa_\rho$ for simplicity,
unless specified otherwise.

\subsection{Bulk graviton and dilaton}

The effective action for the gravitaton  ($\eta_{\mu\nu}\rightarrow\eta_{\mu\nu}+h_{\mu\nu}$) and scalar-dilaton fluctuations ($\phi\rightarrow\phi+\varphi$) follows from the Einstein-Hilbert action plus dilaton in the string frame. In de-Donder gauge and to quadratic order, we have

\be
S=\int d^{5} x \sqrt{g}\,e^{-2\phi}\big(\mathcal{L}_{h+f}+\mathcal{L}_\varphi \big)\,,\nonumber\\
\label{Action2}
\ee
with

\bea
\label{kinetic}
\mathcal{L}_{h+f} =&& -\frac{1}{4\tilde{g}_5^2}\,g^{\mu\nu}\,\eta^{\lambda\rho}\eta^{\sigma\tau}\partial_{\mu}h_{\lambda\sigma}\partial_{\nu}h_{\rho\tau}\nonumber\\
&&+\frac{1}{8\tilde{g}_5^2}\,g^{\mu\nu}\eta^{\alpha\beta}\eta^{\gamma\sigma}\,\partial_{\mu}h_{\alpha\beta}\,\partial_{\nu}h_{\gamma\sigma}\,,\nonumber\\
\mathcal{L}_\varphi=&&+\frac{1}{2\tilde{g}_5^2}\,g^{\mu\nu}\,\partial_{\mu}\varphi\,\partial_{\nu}\varphi\,,
\ee
and $\tilde{g}_5^2=2\kappa^2=16\pi G_N={8\pi^2}/{N_c^2}$. 
The graviton  in bulk is dual to a glueball on the boundary.
We follow~\cite{Kanitscheider:2008kd}, and split $h_{\mu\nu}$ into 
a traceless part $h$ (tensor $2^{++}$ glueball) and and trace-full part $f$ (scalar $0^{++}$ glueball)

\begin{widetext}
\be
\label{exp}
h_{\mu\nu}(k,z)=\bigg[\epsilon_{\mu\nu}^{TT}h(k,z)+k_{\mu} k_{\nu}H(k,z)\bigg]+\bigg[k_{\mu}A^{\perp}_{\nu}(k,z)+k_{\nu}A^{\perp}_{\mu}(k,z)\bigg]+\bigg[\frac{1}{3}\eta_{\mu\nu}f(k,z)\bigg]
\ee
\end{widetext}
with $k^\mu\epsilon_{\mu\nu}^{TT}=\eta^{\mu\nu}\epsilon_{\mu\nu}^{TT}=0$.
A further gauge fixing $A_{\mu}^{\perp}=0$, allows to decouple the tensor glueball $h$.
In contrast, the equations for  $f$, $H$, and $\varphi$ (denoted as $k$ in \cite{Kanitscheider:2008kd}) are coupled
(see Eqs.7.16-20 in \cite{Kanitscheider:2008kd}). Diagonalizing the equations, one can show that $f$ satisfies the same equation of motion as $h$ \cite{Kanitscheider:2008kd}. Also note that $f_{0}=f(z=0)$ couples to $T^{\mu}_{\mu}$ of the gauge theory, while $H_{0}=H(z=0)$ couples to $k^{\mu}k^{\nu}T_{\mu\nu}\equiv 0$ (see Eq.7.6 of \cite{Kanitscheider:2008kd}).

The ensuing spectra for the tensor $2^{++}$ and scalar $0^{++}$ glueballs are degenerate~\cite{HILMAR}

\begin{widetext}
\bea
\label{SOFTMFG2}
m_{T,S}^2(n)=8\tilde\kappa^2_N\bigg(n+2\bigg)\qquad
(T: \tilde{g}_5f_n,\,\,\, S: \sqrt{2}\tilde{g}_5f_n) \rightarrow 2\tilde\kappa_N
\eea
\end{widetext}
They differ from their vector meson counterparts in (\ref{SOFTMF}) by the replacements
 $\tilde\kappa_V\rightarrow\sqrt{2}\tilde\kappa_N$ and $g_5\rightarrow\tilde g_5$ due to the difference in the bulk actions.
The spectrum of the scalar-dilaton fluctuations and coupling are similar to the tensor glueball.

The bulk graviton couplings are

  \be
 h\overline\Psi\Psi:\quad &&-\frac{\sqrt{2\kappa^2}}{2}\int d^5x\,\sqrt{g}\,h_{\mu\nu}T_F^{\mu\nu}\nonumber\\
  h AA:\quad && -\frac{\sqrt{2\kappa^2}}{2}\int d^5x\,\sqrt{g}\,h_{\mu\nu}T_V^{\mu\nu}\nonumber\\
  \label{vertices1}
 \ee
with the energy-momentum tensors

 \bea
T_F^{\mu\nu}&=&e^{-\phi}\frac{i}{2}\,z\,\overline\Psi\gamma^\mu\overset{\leftrightarrow}{\partial^\nu}\Psi-\eta^{\mu\nu}\mathcal{L}_F\,,\nonumber\\
T_V^{\mu\nu} &=&-e^{-\phi}\Big(z^4\eta^{\rho\sigma}\eta^{\mu\beta}\eta^{\nu\gamma}\,F^V_{\beta\rho}F^V_{\gamma\sigma}\nonumber\\
&&-z^4\,\eta^{\mu\beta}\eta^{\nu\gamma}\,F^V_{\beta z}F^V_{\gamma z}\Big)-\eta^{\mu\nu}\mathcal{L}_V\,.\nonumber\\
  \label{EMT}
 \eea
There is no contribution from the  UV-boundary term in (\ref{Action})  since it vanishes for the normalizable modes of the fermion.
The scalar-dilaton couplings are

  \begin{widetext}
  \bea
  \label{vertices2}
 \varphi \bar\Psi\Psi:\quad&& \sqrt{2\kappa^2}\int d^5x\,\sqrt{g}\,\frac{e^{-\phi}}2 \,\left(\frac z2 \,\partial_z\varphi\right)\,
 \overline\Psi\gamma^5\Psi+\sqrt{2\kappa^2}\int d^5x\,\sqrt{g}\,\frac{e^{-\phi}}{2}\,\left(\frac {iz}2 \,\partial_\mu\varphi\right)\,
 \overline\Psi\gamma^\mu\Psi\nonumber\\
\varphi AA:\quad && \sqrt{2\kappa^2}\int d^5x\,\sqrt{g}\,e^{-\phi}\,(-\varphi)\,
 \left(-\frac 1{4}\,g^{\mu\alpha}g^{\beta\nu}\,F_{\mu\nu}^V\,F_{\alpha\beta}^V\right)
\eea
To make the power counting manifest in the Witten diagrams, we canonically rescale the fields as

\bea
\label{SUBX}
\Psi\rightarrow g_5\Psi\qquad V_{N}\rightarrow g_5V_{N}\qquad \varphi\rightarrow\sqrt{2\kappa^2}\,\varphi\qquad h_{\mu\nu}\rightarrow\sqrt{2\kappa^2}\,h_{\mu\nu}
\eea
\end{widetext}
which makes the couplings  and power counting manifest.Note that after this rescaling, the meson decay constants
in (\ref{SOFTMF}) and the glueball decay constants in (\ref{SOFTMFG2}) redefine through  $g_5f_n\rightarrow f_n$. 
This is understood in most of our analysis.

\section{wavefunctions in  holographic QCD}

\subsection{Dirac fermion/proton}

The normalized wavefunctions for the bulk Dirac fermion are~\cite{CARLSON}

\bea
\Psi(p,z)&=&\psi_R(z)\Psi^0_{R}(p)+ \psi_L(z)\Psi^0_{L}(p)\,,\nonumber\\
\bar\Psi(p,z)&=&\psi_R(z)\bar\Psi^0_{R}(p)+ \psi_L(z)\bar\Psi^0_{L}(p)\,,\nonumber\\
\eea
with for the  soft-wall ($\tau=3$)

\be
&&\psi_R(z)=\frac{\tilde{n}_R}{\tilde{\kappa}_{N}^{\tau-2}} z^{\frac{5}{2}}\xi^{\frac{\tau-2}{2}}L_0^{(\tau-2)}(\xi)\,,\nonumber\\
&&\psi_L(z)=\frac{\tilde{n}_L}{\tilde{\kappa}_{N}^{\tau-1}} z^{\frac{5}{2}}\xi^{\frac{\tau-1}{2}}L_0^{(\tau-1)}(\xi)\,,\nonumber\\
\ee
 $L_n^{(\alpha)}(\xi)$, $\tilde{n}_R=\tilde{n}_L \tilde{\kappa}_{N}^{-1}\sqrt{\tau-1}$ are the generalized Laguerre polynomials, 
 and $\tilde{n}_L=\tilde{\kappa}_{N}^{\tau}\sqrt{{2}/{\Gamma(\tau)}}$. The bulk wave functions are normalized 
 
 \be
\int_{0}^{\infty} dz\,\sqrt{g}\,e^{-\phi}\,e^{\mu}_{a}\,\psi_{R/L}^2(z)=\delta^{\mu}_a\,,\nonumber\\
\ee
with $\phi=\tilde{\kappa}_{N}^2z^2$, and the inverse vielbein $e^{\mu}_{a}=\sqrt{\abs{g^{\mu\mu}}}\delta^{\mu}_a$ (no summation intended in $\mu$).
$\Psi^0_{R/L}(p)= P_{\pm}u(p)$, $\bar\Psi^0_{R/L}(p)=\bar u(p)P_{\mp}$, and $P_{\pm}=(1/2)(1\pm \gamma^5)$. The boundary spinors are normalized as

\be
&&\bar u(p)u(p)=2m_N\,,\nonumber\\
&&2m_N\times\bar u(p')\gamma^{\mu}u(p)=\bar u(p')(p'+p)^{\mu}u(p)\,.\nonumber\\
\ee

\subsection{Photon/spin-1 mesons}

The vector wavefunctions are given by~\cite{Grigoryan:2007my}

\be
\phi_n(z)=c_{n}\tilde{\kappa}_V^2z^2 L_n^1( \tilde{\kappa}_V^2z^2)\equiv J_{A}(m_n,z)  \,,
\ee
with $c_{n}=\sqrt{{2}/{n+1}}$ which is determined from the normalization condition (for the soft-wall model with background dilaton $\phi=\tilde{\kappa}_V^2z^2$)
\be
\int dz\,\sqrt{g}e^{-\phi}\,(g^{xx})^2\,\phi_n(z)\phi_m(z)=\delta_{nm}\,.\nonumber\\
\ee
Therefore, we have

\be
F_n=\frac 1{g_5}\bigg(-e^{-\phi}\frac{1}{z^\prime}\partial_{z^\prime}\phi_n(z^\prime)\bigg)_{z^\prime=\epsilon}=-\frac{2}{g_5}c_n(n+1)\tilde{\kappa}_V^2\,,\nonumber\\
\ee
with $\phi_n(z\rightarrow 0)\approx c_n\tilde{\kappa}_V^2z^2(n+1)$. If we define the decay constant as $f_n=-{F_n}/{m_n}$,  we have

\be
\phi_n(z)=\frac{f_n}{m_n}\times 2g_{5}\tilde{\kappa}_V^2z^2 L_n^1( \tilde{\kappa}_V^2z^2)\,,
\ee
as required by vector meson dominance (VMD).

The bulk-to-bulk propagator is

\begin{widetext}
\be
G(z\rightarrow 0,z')&&\approx \frac{\phi_n(z\rightarrow 0)}{-g_5F_n}\sum_n \frac{-g_5F_n\phi_n(z')}{q^2-m_n^2}=\frac{z^2}{2}\sum_n \frac{-g_5F_n\phi_n(z')}{q^2-m_n^2}=\frac{z^2}{2}V(q,z') . \nonumber\\\label{vbbt2sw}
\ee
\end{widetext}
For space-like momenta ($q^2=-Q^2$), we have the bulk-to-bulk propagator near the boundary

\be
G(z\rightarrow 0,z')\approx \frac{z^2}{2}\sum_n \frac{g_5F_n\phi_n(z')}{Q^2+m_n^2}=\frac{z^2}{2}\mathcal{V}(Q,z')\,, \nonumber\\\label{vbbt2sw}
\ee
with~\cite{Grigoryan:2007my}

\bea
\mathcal{V}(Q,z)
=\kappa_V^2z^2\int_{0}^{1}\frac{dx}{(1-x)^2}x^a{\rm exp}\Big[-\frac{x}{1-x}\kappa_V^2z^2\Big]\,,\nonumber\\
\label{vps2sw}
\ee
and  the normalization ${\cal V}(0,z)={\cal V}(Q,0)=1$.

\subsection{Tansverse-traceless graviton/spin-2 glueballs}

Similar relationships hold for the soft-wall model where the normalized wave function for spin-2 glueballs is given by \cite{BallonBayona:2007qr} (note that the discussion in \cite{BallonBayona:2007qr} is for general massive bulk scalar fluctuation but can be used for spin-2 glueball which has an effective bulk action similar to massless bulk scalar fluctuation)
\bea
 \label{wfSW}
&&\psi_{n}(z)=c_n\,z^{4}L_{n}^{\Delta(j)-2}(2\xi)\,,\nonumber\\
\eea
with
\be
c_n=\Bigg(\frac{2^{4}\tilde{\kappa}_{N}^{6}\Gamma(n+1)}{\Gamma(n+3)}\Bigg)^{\frac 12}\,,
\ee
which is determined from the normalization condition (for soft-wall model with background dilaton $\phi=\tilde{\kappa}_N^2z^2$)
\be
\int dz\,\sqrt{g}e^{-\phi}\,\abs{g^{xx}}\,\psi_n(z)\psi_m(z)=\delta_{nm}\,.\nonumber\\
\ee
Therefore we have

\be
F_n=\frac{1}{\sqrt{2}\kappa}\bigg(-\frac{1}{z^{\prime 3}}\partial_{z^\prime}\psi_n(z^\prime)\bigg)_{z^\prime=\epsilon}=-\frac{4}{\sqrt{2}\kappa}c_n L_{n}^{2}(0)\,,\nonumber\\
\ee
with $\psi_n(z\rightarrow 0)\approx c_n\,z^{4}L_{n}^{2}(0)$. 
For space-like momenta ($q^2=-Q^2$), we have the bulk-to-bulk propagator near the boundary
\be
G(z\rightarrow 0,z')\approx \frac{z^4}{4}\sum_n \frac{\sqrt{2}\kappa F_n\phi_n(z')}{K^2+m_n^2}=\frac{z^4}{4}\mathcal{H}(K,z') , \nonumber\\\label{hbbt3SW2}
\ee
where, for the soft-wall model,~\cite{CARLSON,Abidin:2008ku,BallonBayona:2007qr}

\begin{widetext}
\bea
\mathcal{H}(K,z)
&&=4z^{4}\Gamma(a_K +2)U\Big(a_K+2,3;2\xi\Big)
=\Gamma(a_K+2)U\Big(a_K,-1;2\xi\Big)\nonumber\\
&&=\frac{\Gamma(a_K+2)}{\Gamma(a_K)}
\int_{0}^{1}dx\,x^{a_K-1}(1-x){\rm exp}\Big(-\frac{x}{1-x}(2\xi)\Big)\,,
\label{BBSWj2}
\eea
\end{widetext}
with $a_K={a}/{2}={K^2}/{8\tilde{\kappa}_N^2}$, and we have used the transformation $U(m,n;y)=y^{1-n}U(1+m-n,2-n,y)$. (\ref{BBSWj2}) satisfies the normalization condition ${\cal H}(0,z)={\cal H}(K,0)=1$.

\section{Details of the holographic differential cross section: near threshold}\label{dcnt}

The holographic differential cross section  for untraced in and out polarizations
read
\begin{widetext}
\be
\label{DCTT}
&&\frac{d\sigma(s,t,Q,M_{J/\Psi},\epsilon_{T},\epsilon'_{T})}{dt}
=\frac{e^2}{16\pi(s-(-Q^2+m_N^2))^2}\,
\frac 12\sum_{{\rm spin}}
\Bigg|{\cal A}^{TT}_{\gamma^* p\rightarrow  J/\Psi p} (s,t,Q,M_{J/\Psi},\epsilon_{T},\epsilon '_{T})\Bigg|^2\,,\nonumber\\
&&\frac{d\sigma(s,t,Q,M_{J/\Psi},\epsilon_{L},\epsilon'_{L})}{dt}
=\frac{e^2}{16\pi(s-(-Q^2+m_N^2))^2}\,
\frac 12\sum_{{\rm spin}}
\Bigg|{\cal A}^{LL}_{\gamma^* p\rightarrow  J/\Psi p} (s,t,Q,M_{J/\Psi},\epsilon_{L},\epsilon '_{L})\Bigg|^2\,,
\ee
where the transverse and longitudinal amplitudes are respectively

\bea \label{TTamp}
&&{\cal A}^{TT}_{\gamma^* p\rightarrow  J/\Psi p} (s,t,Q,M_{J/\Psi},\epsilon_{T},\epsilon^\prime_{T})=\nonumber\\
&&\frac{1}{g_5}\times\Big(\mathcal{I}(Q,M_{J/\Psi})\times B_{1}^{TT}-\mathcal{J}(Q,M_{J/\Psi})\times B_{0}^{TT}Q^2\Big)\times 2\kappa^2\times A(K)\times\frac{1}{m_N}
\,\bar u(p_2)u(p_1)\,,\nonumber\\
&&{\cal A}^{LL}_{\gamma^* p\rightarrow  J/\Psi p} (s,t,Q,M_{J/\Psi},\epsilon_{L},\epsilon^\prime_{L})=\nonumber\\
&&\frac{1}{g_5}\times\Big(\mathcal{I}(Q,M_{J/\Psi})\times B_{1}^{LL}-\mathcal{J}(Q,M_{J/\Psi})\times B_{0}^{LL}Q^2\Big)\times 2\kappa^2\times A(K)\times\frac{1}{m_N}
\,\bar u(p_2)u(p_1)\,,\nonumber\\
\eea
with
\bea
B_1^{TT}(s,t,Q,M_{J/\Psi})&=&p_\alpha p_\beta B_{1}^{\alpha\beta}(\epsilon_{T},\epsilon '_{T})=\epsilon_{T}\cdot\epsilon '_{T}\,q\cdot p\,q'\cdot p+q\cdot q'\,\epsilon_{T}\cdot p\,\, \epsilon'_{T}\cdot p-q\cdot\epsilon '_{T}\,p\cdot\epsilon_{T}\,p\cdot q'-q'\cdot\epsilon_{T}\,p\cdot\epsilon '_{T}\,p\cdot q \,,\nonumber\\
B_1^{LL}(s,t,Q,M_{J/\Psi})&=&p_\alpha p_\beta B_{1}^{\alpha\beta}(\epsilon_{L},\epsilon '_{L})=\epsilon_{L}\cdot\epsilon '_{L}\,q\cdot p\,q'\cdot p+q\cdot q'\,\epsilon_{L}\cdot p\,\, \epsilon'_{L}\cdot p-q\cdot\epsilon '_{L}\,p\cdot\epsilon_{L}\,p\cdot q'-q'\cdot\epsilon_{L}\,p\cdot\epsilon '_{L}\,p\cdot q \,,\nonumber\\
B_0^{TT}(s,t,Q,M_{J/\Psi})&=&p_\alpha p_\beta B_{0}^{\alpha\beta}(\epsilon_{T},\epsilon '_{T})=\epsilon_{T}\cdot p\,\, \epsilon'_{T}\cdot p\,,\nonumber\\
B_0^{LL}(s,t,Q,M_{J/\Psi})&=&p_\alpha p_\beta B_{0}^{\alpha\beta}(\epsilon_{L},\epsilon '_{L})=\epsilon_{L}\cdot p\,\, \epsilon'_{L}\cdot p\,,
\eea
where we will make use of the orthogonality conditions
\be
\epsilon_{L}\cdot q=\epsilon_{L}^0q_0-\vert\vec\epsilon_{L}\vert q_z \cos\theta' =0
\ee
and 
\be
\epsilon_{L}'\cdot q'=\epsilon_{L}'^0q'_0-\vert\vec\epsilon_{L}'\vert \vert \mathbf{q}_{V}\vert \cos\theta'' =0
\ee
in order to evaluate
\be
\epsilon_{L}\cdot q'=\epsilon_{L}^0q'_0-\vert\vec\epsilon_{L}\vert \vert \mathbf{q}_{V}\vert \cos(\theta'+\theta)\,,
\ee
and 
\be
\epsilon_{L}'\cdot q=\epsilon_{L}'^0q_0-\vert\vec\epsilon_{L}'\vert q_z \cos(\theta''+\theta)\,.
\ee
Also note that we will use $\epsilon_L^2=-1$ and $\epsilon{^\prime 2}_L=-1$ in order to find
\bea
\vert\vec\epsilon_{L}\vert=\sqrt{\frac{Q^2+q_z^2}{Q^2}}\,,\quad\quad\quad\vert\vec\epsilon_{L}'\vert=\sqrt{\frac{M_V^2+\vert \mathbf{q}_{V}\vert^2}{M_V^2}}\,.
\eea
For the transverse part we use
\be
\epsilon_{T}\cdot\epsilon_{T}'=-\vert\vec\epsilon_{T}\vert\vert\vec\epsilon'_{T}\vert\cos\theta=-\cos\theta\,,
\ee
\be
\epsilon_{T}\cdot q'=-\vert\vec\epsilon_{T}\vert \vert \mathbf{q}_{V}\vert \cos\left(\frac{\pi}{2}+\theta\right)=-\vert \mathbf{q}_{V}\vert\sin\theta\,,
\ee
and 
\be
\epsilon_{T}'\cdot q=-\vert\vec\epsilon_{T}'\vert q_z \cos\left(\frac{\pi}{2}-\theta\right)=-q_z \sin(\theta)\,.
\ee
In addition, we have defined
\bea
 \label{IJ}
\mathcal{I}(Q,M_{J/\Psi})
=&&\Big(\frac{\tilde{\kappa}_{J/\Psi}}{Q}\Big)^4\times\frac{1}{2}\int_{0}^{\infty} d\xi\,e^{-\xi^2\frac{\tilde{\kappa}_{J/\Psi}^2}{Q^2}}\,\xi^{-1}\times \mathcal{V}_{\gamma^*}(\xi)\mathcal{V}_{J/\Psi}(\xi M_{J/\Psi}/Q)\times \frac{\xi^{4}}{4}\nonumber\\
=&&\frac{f_{J/\Psi}}{M_{J/\Psi}}\times g_5\times\Bigg(\frac{3}{\frac{1}{32}\frac{Q^6}{\tilde{\kappa}_{J/\Psi}^6}+\frac{3}{4}\frac{Q^4}{\tilde{\kappa}_{J/\Psi}^4}+\frac{11}{2}\frac{Q^2}{\tilde{\kappa}_{J/\Psi}^2}+12}\Bigg)\nonumber\\
=&&\frac{1}{2}\frac{f_{J/\Psi}}{M_{J/\Psi}}\times g_5\times\left(\frac{3}{\left(\frac{Q^2}{4 \tilde{\kappa}_{J/\Psi}^2}+3\right)\left(\frac{Q^2}{4 \tilde{\kappa}_{J/\Psi}^2}+2\right)\left(\frac{Q^2}{4 \tilde{\kappa}_{J/\Psi}^2}+1\right)}\right)
\,,\nonumber\\
\mathcal{J}(Q,M_{J/\Psi})
=&&\Big(\frac{\tilde{\kappa}_{J/\Psi}}{Q}\Big)^4\times\frac{1}{2}\int_{0}^{\infty} d\xi\,e^{-\xi^2\frac{\tilde{\kappa}_{J/\Psi}^2}{Q^2}}\,\xi^{-1}\times\partial_{\xi}\mathcal{V}_{\gamma^*}(\xi)\times\partial_{\xi}\mathcal{V}_{J/\Psi}(\xi M_{J/\Psi}/Q)\times \frac{\xi^{4}}{4}\nonumber\\
=&&-\frac{f_{J/\Psi}}{M_{J/\Psi}}\times g_5\times\Bigg(\frac{1}{\frac{1}{32}\frac{Q^6}{\tilde{\kappa}_{J/\Psi}^6}+\frac{3}{4}\frac{Q^4}{\tilde{\kappa}_{J/\Psi}^4}+\frac{11}{2}\frac{Q^2}{\tilde{\kappa}_{J/\Psi}^6}+12}\Bigg)\nonumber\\
=&&-\frac{1}{3}\times\mathcal{I}(Q,M_{J/\Psi})\,,
\eea
with $\xi\equiv Qz$. 

Following the general invariant decomposition

\be
\label{EMT2}
\left<p_2|T^{\mu\nu}(0)|p_1\right>=\overline{u}(p_2)\left(
A(k)\gamma^{(\mu}p^{\nu)}+B(k)\frac{ip^{(\mu}\sigma^{\nu)\alpha}k_\alpha}{2m_N}+C(k)\frac{k^\mu k^\nu-\eta^{\mu\nu}k^2}{m_N}\right)u(p_1)\,,
\ee
the tensor gravitational form factor is

\be\label{Aff}
A=\frac{1}{2}\int dz\sqrt{g}\,e^{-\phi}z\,\big(\psi_R^2(z)+\psi_L^2(z)\big)\,\mathcal{H}(K,z)
\ee
and   $B(0)=0$, in the present holographic construction.  The trace is normalized  $\left<p|T^{\mu}_\mu|p\right>=2A(0)m_N^2$. Specifically, for  the soft wall model, we have

\be \label{FFj2}
A(K)=A(0) \bigg((1-2a_K)(1+a_K^2)+a_K(1+a_K)(1+2a_K^2)\bigg(H\bigg(\frac{1+a_K}{2} \bigg)-H\bigg(\frac{a_K}{2}\bigg)\bigg)\bigg)
\ee
with $a_K={K^2}/{8\tilde\kappa_N^2}$. Here $H(x)$ is the harmonic number $H(x)=\psi(1+x)+\gamma$. 
The gravitational form factor $B(K)$  is found to be null, and the gravitational form factor $C(K)$ is fixed
by both the tensor and scalar glueball contributions. It will not be needed here.
Since the boundary value ${\cal H}(K,0)={\cal H}(0,z)$ is arbitrary (1-point function), it follows that
$A(0)$ is not fixed in holography.

Finally, evaluating the spin sum over the initial and final bulk Dirac fermions, we find

\bea
\label{DCTT2}
\frac{d\sigma(s,t,Q,M_{J/\Psi},\epsilon_{T},\epsilon'_{T})}{dt}
&=&\frac{e^2\times\frac{(2\kappa^2)^2}{g_5^2}}{16\pi(s-(-Q^2+m_N^2))^2}\,
\frac 12\times \left(\frac{s}{\tilde{\kappa}_N^2}\right)^{4}
\times  \mathcal{I}^2(Q,M_{J/\Psi})\times\frac{\tilde{\kappa}_N^8}{\tilde{\kappa}_{J/\Psi}^8}\nonumber\\
&&\times\frac{1}{s^4}\tilde{F}^{TT}(s,t,Q,M_{J/\Psi},m_N)\times(2K^2+8m_N^2)\times\frac{1}{m_N^2}\times A^2(t)\,,\nonumber\\
\frac{d\sigma(s,t,Q,M_{J/\Psi},\epsilon_{L},\epsilon'_{L})}{dt}
&=&\frac{e^2\times\frac{(2\kappa^2)^2}{g_5^2}}{16\pi(s-(-Q^2+m_N^2))^2}\,
\frac 12\times \left(\frac{s}{\tilde{\kappa}_N^2}\right)^{4}
\times \mathcal{I}^2(Q,M_{J/\Psi})\times\frac{\tilde{\kappa}_N^8}{\tilde{\kappa}_{J/\Psi}^8}\nonumber\\
&&\times\frac{1}{s^4}\tilde{F}^{LL}(s,t,Q,M_{J/\Psi},m_N)\times(2K^2+8m_N^2)\times\frac{1}{m_N^2}\times A^2(t)
\eea
where we have used the spin sum
\bea
\sum_{s,s'}\bar u_{s'}(p_2)u_s(p_1)\bar u_s(p_1)u_{s'}(p_2)&=& \Tr\Big(\sum_{s.s'}u_{s'}(p_2)\bar u_{s'}(p_2)u_s(p_1)\bar u_s(p_1)\Big)\nonumber\\
&=&\frac 14 \Tr\Big(\big(\gamma_\mu p_2^\mu+m_N\big)\big(\gamma_\mu p_1^\mu+m_N\big)\Big)=2K^2+8m_N^2\,,
\eea
and defined the kinematic factors coming from the polarization tensors as
\bea
\label{FTTFLL}
\tilde{F}^{TT}(s,t,Q,M_{J/\Psi},m_N)&=&\Big[B_1^{TT}(s,t,Q,M_{J/\Psi})\Big]^2+\frac{1}{9}\times\Big[B_0^{TT}(s,Q,M_{J/\Psi})Q^2\Big]^2\nonumber\\
&-&2\times \frac{1}{3}\times B_1^{TT}(s,t,Q,M_{J/\Psi})\times B_0^{TT}(s,t,Q,M_{J/\Psi})Q^2\,,\nonumber\\
\tilde{F}^{LL}(s,t,Q,M_{J/\Psi},m_N)&=&\Big[B_1^{LL}(s,t,Q,M_{J/\Psi})\Big]^2+\frac{1}{9}\times\Big[B_0^{LL}(s,t,Q,M_{J/\Psi})Q^2\Big]^2\nonumber\\
&-&2\times \frac{1}{3}\times B_1^{LL}(s,t,Q,M_{J/\Psi})\times B_0^{LL}(s,t,Q,M_{J/\Psi})Q^2\,.
\eea

\begin{figure}[!htb]
\minipage{0.8\textwidth}
  \includegraphics[width=8cm]{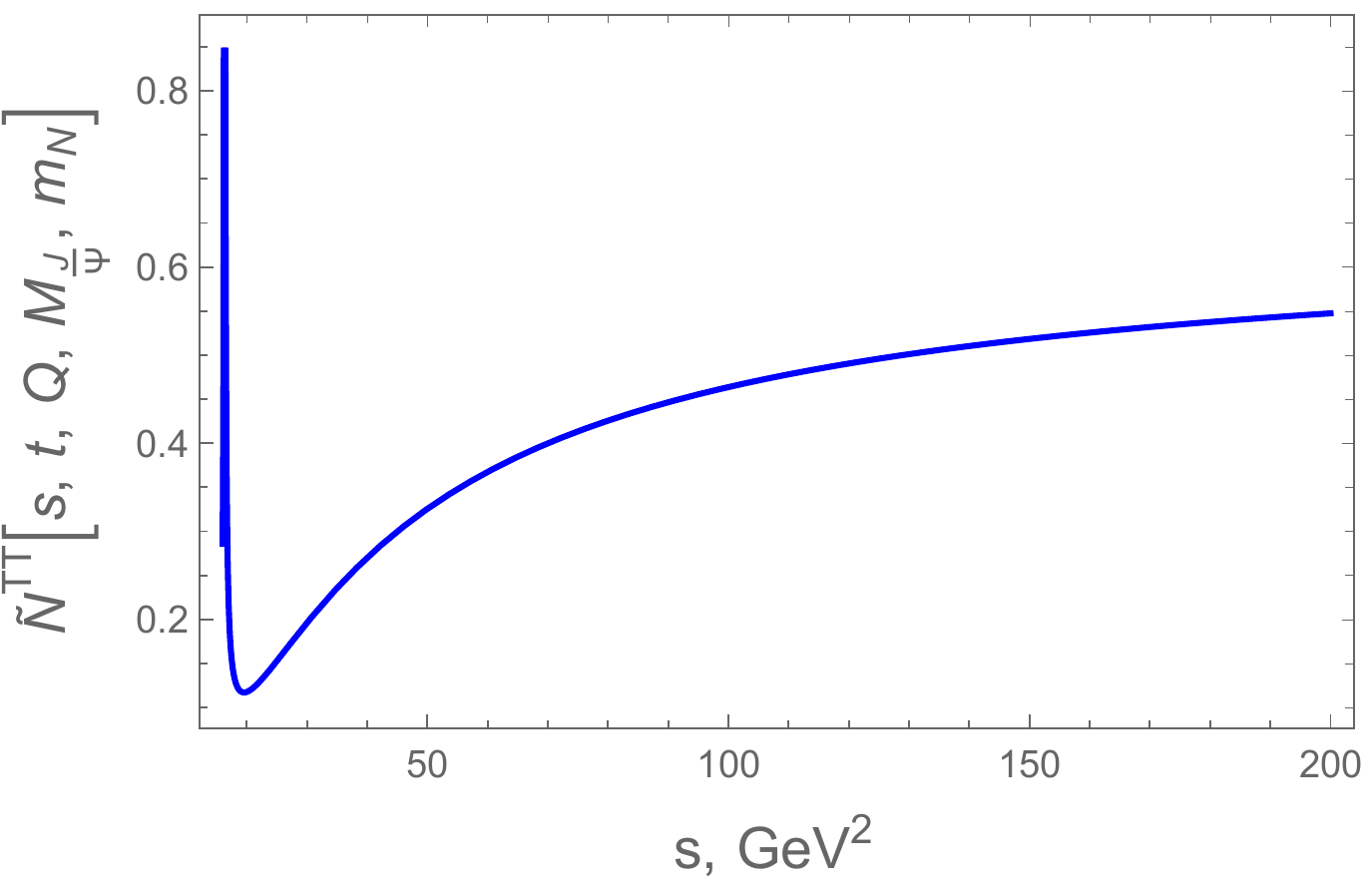}
\endminipage\hfill
\minipage{0.8\textwidth}
 \includegraphics[width=8cm]{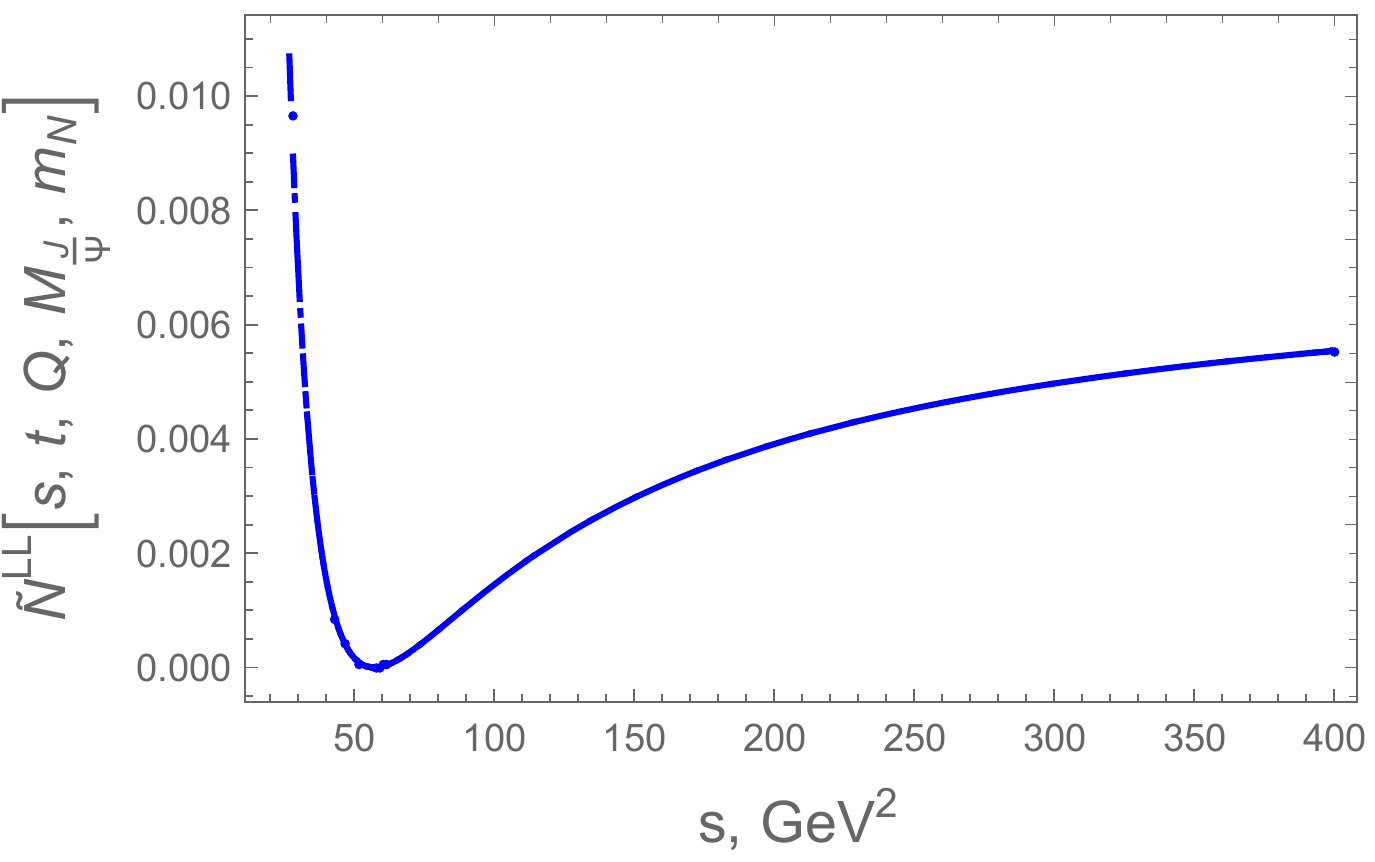}
\endminipage\hfill
  \caption{The rescaled transverse normalization coefficient in (\ref{NTTLL})  (top),  and  rescaled longitudinal normalization coefficient in (\ref{NTTLL}) (bottom),
 for  $t=-1~{\rm GeV}^2$, $Q=1~{\rm GeV}$, $M_{J/\Psi}=3.10~{\rm GeV}$ and $m_N=0.94~{\rm GeV}$.}
 \label{fig_NTTNLL}
\end{figure}

We can further rewrite the differential cross sections more compactly as
\bea
\label{DCTT2}
\frac{d\sigma(s,t,Q,M_{J/\Psi},\epsilon_{T},\epsilon'_{T})}{dt}
&=&\mathcal{I}^2(Q,M_{J/\Psi})\times\left(\frac{s}{\tilde{\kappa}_N^2}\right)^{2}\times\mathcal{N}^{TT}(s,t,Q,M_{J/\Psi},m_N)\times\left(-\frac{t}{4m_N^2}+1\right)\times\tilde{A}^2(t)\,,\nonumber\\
\frac{d\sigma(s,t,Q,M_{J/\Psi},\epsilon_{L},\epsilon'_{L})}{dt}
&=&\mathcal{I}^2(Q,M_{J/\Psi})\times\left(\frac{s}{\tilde{\kappa}_N^2}\right)^{2}\times\frac{1}{9}\times\frac{Q^2}{M_{J/\Psi}^2}\times\mathcal{N}^{LL}(s,t,Q,M_{J/\Psi},m_N)\times\left(-\frac{t}{4m_N^2}+1\right)\times \tilde{A}^2(t)\,,\nonumber\\
\eea
where we have defined the normalization coefficients as
\bea \label{NTTLL}
\mathcal{N}^{TT}(s,t,Q,M_{J/\Psi},m_N)&=&\frac{e^2\times\frac{(2\kappa^2)^2}{g_5^2}}{16\pi(1-(-\frac{Q^2}{s}+\frac{m_N^2}{s}))^2}\,
\frac 12\times\frac{\tilde{\kappa}_N^4}{\tilde{\kappa}_{J/\Psi}^8}\times A^2(0)\nonumber\\
&\times&\frac{1}{s^4}\tilde{F}^{TT}(s,t,Q,M_{J/\Psi},m_N)\times 8\,,\nonumber\\
\mathcal{N}^{LL}(s,t,Q,M_{J/\Psi},m_N)&=&\frac{e^2\times\frac{(2\kappa^2)^2}{g_5^2}}{16\pi(1-(-\frac{Q^2}{s}+\frac{m_N^2}{s}))^2}\times\frac 12\times\frac{\tilde{\kappa}_N^4}{\tilde{\kappa}_{J/\Psi}^8}\times A^2(0)\,\nonumber\\
&\times & 9\times\frac{M_{J/\Psi}^2}{Q^2}\times\frac{1}{s^4}\tilde{F}^{LL}(s,t,Q,M_{J/\Psi},m_N)\times 8\,\nonumber\\
\eea
and we have normalized the gravitational form factor $A(t)$ to be unity at $t=0$ as
\be
\tilde{A}(t)\equiv\frac{A(t)}{A(0)}\,.
\ee

The $TT$ and $LL$ kinematical functions $\tilde F$ in (\ref{NTTLL}) are given in (\ref{FTTFLL}). 
In  Figs.~\ref{fig_NTTNLL} we show the behavior of the rescaled normalizations (\ref{NTTLL}) over a broad range of $s$. The rescaling is through
${\cal N}\rightarrow \tilde{\cal N}={\cal N}/f$ with  all holographic couplings lumped in 
\be
f\equiv f_{TT,LL}=\Bigg[{e^2\frac{(2\kappa^2)^2}{32\pi g_5^2}\frac{\tilde{\kappa}_N^4}{\tilde{\kappa}_{J/\Psi}^8}}\times\frac{1}{\left(-\frac{t}{4m_N^2}+1\right)}~,~{e^2\frac{(2\kappa^2)^2}{32\pi g_5^2}\frac{\tilde{\kappa}_N^4}{\tilde{\kappa}_{J/\Psi}^8}}\times 9\times\frac{M_{J/\Psi}^2}{Q^2}\times\frac{1}{\left(-\frac{t}{4m_N^2}+1\right)}\Bigg]\,.\nonumber\\
\ee

For heavy vector mesons, like $J/\Psi$, we can assume $s\sim s_{threshold}\gg m_N^2$, and on top of that if we restrict ourselves to $s\sim s_{threshold}\gg Q^2, M_{V}^2, |t|$, we have $\epsilon_{L}^{\mu}(s\rightarrow\infty)=\frac{q^\mu}{Q}$, $\epsilon_{L}'^{\mu}(s\rightarrow\infty)=\frac{q'^\mu}{M_V}$, and $\theta(s\rightarrow\infty)=0$ hence $\epsilon_{T}\cdot\epsilon'_{T}(s\rightarrow\infty)=-1$, $\epsilon_{T}\cdot A(s\rightarrow\infty)=\epsilon'_{T}\cdot A(s\rightarrow\infty)=0$ for $A=q,q',p,p_1,p_2$. Therefore, for heavy mesons, we have
\be
\tilde{F}(s)=&&\tilde{F}^{TT}(s\rightarrow\infty,t,Q,M_{J/\Psi},m_N)\nonumber\\
=&&9\times\frac{M_{J/\Psi}^2}{Q^2}\times\tilde{F}^{LL}(s\rightarrow\infty,t,Q,M_{J/\Psi},m_N)\nonumber\\
=&&Q^2M_{J/\Psi}^2\times\Big[B_0^{LL}(s\rightarrow\infty,t,Q,M_{J/\Psi})\Big]^2\nonumber\\
=&&\Big[B_1^{TT}(s\rightarrow\infty,t,Q,M_{J/\Psi})\Big]^2\nonumber\\
=&&\lim_{s \to \infty}(q\cdot p\,q'\cdot p)^2=\lim_{s \to \infty}(2q_z^2)^4=\frac{1}{16}\times s^4\,,
\ee
which simplifies the normalization coefficients to
\bea
\mathcal{N}^{TT}(s,t,Q,M_{J/\Psi},m_N)&\approx&\frac{e^2\times\frac{(2\kappa^2)^2}{g_5^2}}{16\pi}\,
\frac 12\times\frac{\tilde{\kappa}_N^4}{\tilde{\kappa}_{J/\Psi}^8}\times A^2(0)\times\frac{1}{s^4}\tilde{F}(s)\times 8=constant\,,\nonumber\\
\mathcal{N}^{LL}(s,t,Q,M_{J/\Psi},m_N)
&\approx &\mathcal{N}^{TT}(s,t,Q,M_{J/\Psi},m_N)\,,\nonumber\\
\eea
where $\frac{1}{s^4}\tilde{F}(s)=\frac{1}{16}$.

\end{widetext}

\section{Details of the holographic differential cross section: high energy regime}\label{dcher}


In the high energy limit $\sqrt{\lambda}/\tilde{\tau}\rightarrow 0$
with $\tilde{\tau}\equiv\log\tilde{s}=\log[s/\tilde{\kappa}_N^2]$, following our recent analysis of the photoproduction process, the spin j-exchange for the transverse and longitudinal amplitudes reads~\cite{Mamo:2019mka}

\begin{widetext}

\bea \label{pomeron3}
{\cal A}^{TT}_{\gamma^* p\rightarrow  J/\Psi p} (s,t,Q,M_{J/\Psi},\epsilon_{T},\epsilon '_{T})&\simeq & e^{j_0\tilde{\tau}} \left[(\sqrt{\lambda}/\pi)+ i\right] ( \sqrt{\lambda}/ 2 \pi )^{1/2}\; \nonumber\\
&\times& \tilde{\xi}  \; \frac{e^ {-\sqrt\lambda  \tilde{\xi}^2 / 2\tilde{\tau}}}{\tilde{\tau}^{3/2}}\left(1 + {\cal O}\bigg(\frac{\sqrt{\lambda}}{\tilde{\tau}}\bigg) \right)
\times  G_{5}^{TT}(j_0,s,t,Q,M_{J/\Psi})\,,\nonumber\\
{\cal A}^{LL}_{\gamma^* p\rightarrow  J/\Psi p} (s,t,Q,M_{J/\Psi},\epsilon_{T},\epsilon '_{T})&\simeq & e^{j_0\tilde{\tau}} \left[(\sqrt{\lambda}/\pi)+ i\right] ( \sqrt{\lambda}/ 2 \pi )^{1/2}\; \nonumber\\
&\times&\tilde{\xi}  \; \frac{e^ {-\sqrt\lambda  \tilde{\xi}^2 / 2\tilde{\tau}}}{\tilde{\tau}^{3/2}}\left(1 + {\cal O}\bigg(\frac{\sqrt{\lambda}}{\tilde{\tau}}\bigg) \right)
\times  G_{5}^{LL}(j_0,s,t,Q,M_{J/\Psi})
\eea
with $\tilde{\xi}-\pi/2=\gamma=0.55772.....$ is Euler-Mascheroni constant, and
\bea
\label{G5TTLL}
G_{5}^{TT}(j_0,s,t,Q,M_{J/\Psi})&=&\Big(\frac{\tilde{\kappa}_N}{\tilde{\kappa}_V}\Big)^{4-\Delta(j)+j-2}
\times\frac{1}{s^2}\bigg[\frac{1}{2}\tilde{\kappa}_V^{4-\Delta(j)+j-2}\Gamma(\Delta(j)-2)\nonumber\\
&&\times \Big(\mathcal{V}^{\mathcal{I}}_{h\gamma^* J/\Psi}(j,Q,M_{J/\Psi})\times B_{1}^{TT}-\mathcal{V}^{\mathcal{J}}_{h\gamma^* J/\Psi}(j,Q,M_{J/\Psi})\times B_{0}^{TT}\Big)\nonumber\\
&&\times\frac{\sqrt{2\kappa^2}}{g_5}\times \tilde{\kappa}_N^{j-2+\Delta(j)}A(j,K)
 \times\frac{1}{m_N}\times\bar u(p_2)u(p_1)\bigg]_{j\rightarrow j_0,\,\Delta(j)\rightarrow 2}\,,\nonumber\\\nonumber\\
G_{5}^{LL}(j_0,s,t,Q,M_{J/\Psi})&=&\Big(\frac{\tilde{\kappa}_N}{\tilde{\kappa}_V}\Big)^{4-\Delta(j)+j-2}
\times\frac{1}{s^2}\bigg[\frac{1}{2}\tilde{\kappa}_V^{4-\Delta(j)+j-2}\Gamma(\Delta(j)-2)\nonumber\\
&&\times\Big(\mathcal{V}^{\mathcal{I}}_{h\gamma^* J/\Psi}(j,Q,M_{J/\Psi})\times B_{1}^{LL}-\mathcal{V}^{\mathcal{J}}_{h\gamma^* J/\Psi}(j,Q,M_{J/\Psi})\times B_{0}^{LL}\Big)
\nonumber\\\nonumber\\
&&\times\frac{\sqrt{2\kappa^2}}{g_5}\times \tilde{\kappa}_N^{j-2+\Delta(j)}A(j,K)
 \times\frac{1}{m_N}\times\bar u(p_2)u(p_1)\bigg]_{j\rightarrow j_0,\,\Delta(j)\rightarrow 2}
\eea
with, $Qz=\xi$,

\bea
 \label{vvJPSI1j}
\mathcal{V}^{\mathcal{I}}_{h\gamma^* J/\Psi}(j,Q,M_{J/\Psi})=&&\frac{\sqrt{2\kappa^2}}{2}\int_{0}^{\infty} dz\sqrt{g}e^{-z^2\tilde{\kappa}_V^2}\,z^{4+2(j-2)}\nonumber\\
&&\times \mathcal{V}_{\gamma^*}(Q,z)\mathcal{V}_{J/\Psi}(M_{J/\Psi},z)\times C(j) \times z^{\Delta(j)-(j-2)}\nonumber\\
=&& Q^{2-(j+\Delta(j))}\times\frac{\sqrt{2\kappa^2}}{2}\int_{0}^{\infty} d\xi\,e^{-\xi^2\frac{\tilde{\kappa}_V^2}{Q^2}}\,\xi^{4+2(j-2)-5}\nonumber\\
&&\times \mathcal{V}_{\gamma^*}(\xi)\mathcal{V}_{J/\Psi}(\xi M_{J/\Psi}/Q)\times C(j) \times \xi^{\Delta(j)-(j-2)}\,,\nonumber\\\nonumber\\
\mathcal{V}^{\mathcal{J}}_{h\gamma^* J/\Psi}(j,Q,M_{J/\Psi})=&&\frac{\sqrt{2\kappa^2}}{2}\int_{0}^{\infty} dz\sqrt{g}e^{-z^2\tilde{\kappa}_V^2}\,z^{4+2(j-2)}\nonumber\\
&&\times \partial_z\mathcal{V}_{\gamma^*}(Q,z)\times\partial_z\mathcal{V}_{J/\Psi}(M_{J/\Psi},z)\times C(j) \times z^{\Delta(j)-(j-2)}\nonumber\\
=&& Q^{4-(j+\Delta(j))}\times\frac{\sqrt{2\kappa^2}}{2}\int_{0}^{\infty} d\xi\,e^{-\xi^2\frac{\tilde{\kappa}_V^2}{Q^2}}\,\xi^{4+2(j-2)-5}\nonumber\\
&&\times\partial_{\xi}\mathcal{V}_{\gamma^*}(\xi)\times\partial_{\xi}\mathcal{V}_{J/\Psi}(\xi M_{J/\Psi}/Q)\times C(j) \times \xi^{\Delta(j)-(j-2)}\,,
\eea
and

\be\label{Aj3}
&&A(j,K)=\frac{2^{2-\Delta(j)}\tilde{\kappa}_N^{-(j-2)-\Delta(j)}}{4}\times\bigg[\bigg(\frac{\tilde{n}_R}{\tilde{\kappa}_N^{\tau-1}}\bigg)^2\,\Gamma (c) \Gamma (-b+c+1) \, _2F_1(\tilde{a},c;\tilde{a}-b+c+1;-1)\nonumber\\
&+&\bigg(\frac{\tilde{n}_L}{\tilde{\kappa}_N^{\tau}}\bigg)^2\Gamma (c+1) \Gamma (-b+c+2) \, _2F_1(\tilde{a},c+1;\tilde{a}-b+c+2;-1)\bigg]\nonumber\\
&=&\frac{2^{1-\Delta }}{\Gamma (\tau )}\Bigg((\tau -1) \Gamma \left(\frac{j}{2}+\tau -\frac{\Delta }{2}\right) \Gamma \left(\frac{1}{2} (j+\Delta +2 \tau -4)\right) \, _2F_1\left(\frac{1}{2} (j-\Delta +2 \tau ),\frac{1}{2} \left(-\Delta +2a_k+4\right);\frac{1}{2} \left(j+2 \tau +2a_k\right);-1\right)\nonumber\\
&+&\Gamma \left(\frac{j}{2}+\tau -\frac{\Delta }{2}+1\right) \Gamma \left(\frac{1}{2} (j+\Delta +2 \tau -2)\right) \, _2F_1\left(\frac{1}{2} (j-\Delta +2 \tau +2),\frac{1}{2} \left(-\Delta +2a_k+4\right);\frac{1}{2} \left(j+2 \tau +2a_k+2\right);-1\right)\Bigg)\,.\nonumber\\
\ee

The parameters are fixed as
\bea
&&1-\tilde{b}+c=(\tau-1)+\frac{j-2}{2}+\frac{\Delta(j)}{2}\nonumber\\
&&1-\tilde{b}+c+\tilde{a}=(\tau+1)+\frac{j-2}{2}+a_K\nonumber\\
&&c=(\tau+1)+\frac{j-2}{2}-\frac{\Delta(j)}{2}\nonumber\\
&&\tilde{n}_R=\tilde{n}_L \tilde{\kappa}_N^{-1}\sqrt{\tau-1}\qquad
\tilde{n}_L=\tilde{\kappa}_N^{\tau}\sqrt{{2}/{\Gamma(\tau)}}\nonumber\\
\eea
and
\bea
\label{PARA}
&&C(j)= \tilde{\kappa}_V^{2\Delta(j)-4} \times\frac 4{\Delta(j)}
\frac{2^{\Delta(j)-2}\Gamma(a_K+\frac{\Delta(j)}{2})}{\Gamma(\Delta(j)-2)}\qquad\nonumber\\
&&\Delta(j)=2+\sqrt{2\sqrt{\lambda}(j-j_0)}\qquad{\rm and}\qquad a_K=\frac a2=\frac{K^2}{8\tilde{\kappa}_N^2}\qquad{\rm and}\qquad j_0=2-\frac{2}{\sqrt{\lambda}}\,.
\eea

We can rewrite $G_{5}^{TT,LL}(j_0,s,t,Q,M_{J/\Psi})$ more compactly as
\bea
\label{IJ2}
&&G_{5}^{TT}(j_0,s,t,Q,M_{J/\Psi})=\frac{2\kappa^2}{g_5}\times\Big(\frac{\tilde{\kappa}_N}{\tilde{\kappa}_V}\Big)^{4-\Delta(j)+j-2}\times\Big(\frac{Q}{\tilde{\kappa}_V}\Big)^{2-(j+\Delta(j))}\times\frac{4}{\Delta(j)}\times\frac{1}{2}\times\left(\frac{\tilde{\kappa}_V}{\tilde{\kappa}_N}\right)^{2\Delta(j)-4}\times\Big(\frac{\tilde{\kappa}_{J/\Psi}^2}{Q^2}\Big)^{\frac{1}{2} (-\Delta(j)-j+2+4)}\nonumber\\
&&\times\frac{1}{s^2}\bigg(\mathcal{I}(j,Q,M_{J/\Psi})\times B_{1}^{TT}-\mathcal{J}(j,Q,M_{J/\Psi})\times B_{0}^{TT}Q^2\bigg)\times\mathcal{A}(j,\tau,\Delta, K)\times\frac{1}{2m_N}\times\bar u(p_2)u(p_1)\bigg\vert_{j\rightarrow j_0,\,\Delta(j)\rightarrow 2}\,,\nonumber\\ 
&&G_{5}^{LL}(j_0,s,t,Q,M_{J/\Psi})=\frac{2\kappa^2}{g_5}\times\Big(\frac{\tilde{\kappa}_N}{\tilde{\kappa}_V}\Big)^{4-\Delta(j)+j-2}\times\Big(\frac{Q}{\tilde{\kappa}_V}\Big)^{2-(j+\Delta(j))}\times\frac{4}{\Delta(j)}\times\frac{1}{2}\times\left(\frac{\tilde{\kappa}_V}{\tilde{\kappa}_N}\right)^{2\Delta(j)-4}\times\Big(\frac{\tilde{\kappa}_{J/\Psi}^2}{Q^2}\Big)^{\frac{1}{2} (-\Delta(j)-j+2+4)}\nonumber\\
&&\times\frac{1}{s^2}\bigg(\mathcal{I}(j,Q,M_{J/\Psi})\times B_{1}^{LL}-\mathcal{J}(j,Q,M_{J/\Psi})\times B_{0}^{LL}Q^2\bigg)\times\mathcal{A}(j,\tau,\Delta, K)\times\frac{1}{2m_N}\times\bar u(p_2)u(p_1)\bigg\vert_{j\rightarrow j_0,\,\Delta(j)\rightarrow 2}\,,\nonumber\\ 
\eea
and we have defined the dimensionless functions
\bea
\mathcal{A}(j,\tau,\Delta{j}, K) &\equiv &\left(\frac{\tilde{\kappa}_V}{\tilde{\kappa}_N}\right)^{4-2\Delta(j)}\times \tilde{\kappa}_N^{j+2-\Delta(j)}\times\frac{\Delta(j)}{4}\times\Gamma(\Delta(j)-2)\times C(j,K)\times A(j,K)\,,\nonumber\\ \nonumber\\\nonumber\\
&=& 2^{\Delta(j)-2}\times\Gamma\left(a_K+\frac{\Delta(j)}{2}\right)\times\tilde{\kappa}_N^{(j-2)+\Delta(j)}A(j,K)\,,\nonumber\\
\mathcal{I}(j,Q,M_{J/\Psi}) &\equiv & \Big(\frac{\tilde{\kappa}_{J/\Psi}^2}{Q^2}\Big)^{-\frac{1}{2} (-\Delta(j)-j+2+4)}\times\Big(\frac{\tilde{\kappa}_{J/\Psi}}{Q}\Big)^4\times\frac{1}{2}\int_{0}^{\infty} d\xi\,e^{-\xi^2\frac{\tilde{\kappa}_{J/\Psi}^2}{Q^2}}\,\frac{\xi^{\Delta(j)+j+2-5}}{4}\times \mathcal{V}_{\gamma^*}(\xi)\mathcal{V}_{J/\Psi}(\xi M_{J/\Psi}/Q)\nonumber\\
&=&\frac{1}{2}\frac{f_{J/\Psi}}{M_{J/\Psi}}\times g_5\times\frac{\Gamma \left(\frac{Q^2}{4 \tilde{\kappa}_{J/\Psi}^2}+1\right)}{\Gamma\left(\frac{Q^2}{4\tilde{\kappa}_{J/\Psi}^2}+\frac{1}{2}(j+\Delta (j)+2)\right)}\times\left(\frac{j+\Delta(j)}{2}\right)\times\frac{1}{4}\Gamma^2 \left(\frac{j+\Delta(j)}{2}\right)\nonumber\\
&=&\frac{1}{2}\frac{f_{J/\Psi}}{M_{J/\Psi}}\times g_5\times\frac{\Gamma \left(\frac{Q^2}{4 \tilde{\kappa}_{J/\Psi}^2}+1\right)}{\Gamma\left(\frac{Q^2}{4\tilde{\kappa}_{J/\Psi}^2}+\frac{1}{2}(j+\Delta (j))-2\right)}\times\left(\frac{j+\Delta(j)}{2}\right)\times\frac{1}{4}\Gamma^2 \left(\frac{j+\Delta(j)}{2}\right)\nonumber\\
&&\times \frac{1}{\left(\frac{Q^2}{4\tilde{\kappa}_{J/\Psi}^2}+\frac{1}{2}(j+\Delta (j))\right)\left(\frac{Q^2}{4\tilde{\kappa}_{J/\Psi}^2}+\frac{1}{2}(j+\Delta (j))-1\right)\left(\frac{Q^2}{4\tilde{\kappa}_{J/\Psi}^2}+\frac{1}{2}(j+\Delta (j))-2\right)}
\,,\nonumber\\\nonumber\\\nonumber\\
\mathcal{J}(j,Q,M_{J/\Psi}) &\equiv & \Big(\frac{\tilde{\kappa}_{J/\Psi}^2}{Q^2}\Big)^{-\frac{1}{2} (-\Delta(j)-j+2+4)}\times\Big(\frac{\tilde{\kappa}_{J/\Psi}}{Q}\Big)^4\times\frac{1}{2}\int_{0}^{\infty} d\xi\,e^{-\xi^2\frac{\tilde{\kappa}_{J/\Psi}^2}{Q^2}}\,\frac{\xi^{\Delta(j)+j+2-5}}{4}\times\partial_{\xi}\mathcal{V}_{\gamma^*}(\xi)\times\partial_{\xi}\mathcal{V}_{J/\Psi}(\xi M_{J/\Psi}/Q)\,,\nonumber\\
&=&-\frac{1}{2}\frac{f_{J/\Psi}}{M_{J/\Psi}}\times g_5\times\frac{\Gamma \left(\frac{Q^2}{4 \tilde{\kappa}_{J/\Psi}^2}+1\right)}{\Gamma\left(\frac{Q^2}{4\tilde{\kappa}_{J/\Psi}^2}+\frac{1}{2}(j+\Delta (j)+2)\right)}\times\frac{1}{4}\Gamma^2 \left(\frac{j+\Delta(j)}{2}\right)\nonumber\\
&=&-\left(\frac{2}{j+\Delta(j)}\right)\times\mathcal{I}(j,Q,M_{J/\Psi})\,.
\eea

Finally, evaluating the spin sum over the initial and final bulk Dirac fermions, we find, in the high energy regime,
\bea
\label{DCTT2j}
\frac{d\sigma(s,t,Q,M_{J/\Psi},\epsilon_{T},\epsilon'_{T})}{dt}
&=&\frac{e^2\times\frac{(2\kappa^2)^2}{g_5^2}}{16\pi s^2}\,
\frac 12\times\Bigg(\frac{s}{\tilde{\kappa}_N^2}\Bigg)^{4\big(1-\frac{1}{\sqrt{\lambda}}\big)}\times P(\tilde{s},\lambda)\times\Bigg(\frac{\tilde{\kappa}_N^2}{\tilde{\kappa}_{J/\Psi}^2}\Bigg)^{-2\big(1+\frac{1}{\sqrt{\lambda}}\big)}\nonumber\\
&&\times  \mathcal{I}^2(j_0,Q,M_{J/\Psi})\times\frac{\tilde{\kappa}_N^8}{\tilde{\kappa}_{J/\Psi}^8}\times \frac{1}{s^4}\tilde{F}(s)\nonumber\\
&&\times 8\times\mathcal{A}^2(j_0,\tau,\Delta, K)\,,
\eea
and 

\bea
\label{DCLL2j}
\frac{d\sigma(s,t,Q,M_{J/\Psi},\epsilon_{L},\epsilon'_{L})}{dt}
&=&\frac{e^2\times\frac{(2\kappa^2)^2}{g_5^2}}{16\pi s^2}\,
\frac 12\times\Bigg(\frac{s}{\tilde{\kappa}_N^2}\Bigg)^{4\big(1-\frac{1}{\sqrt{\lambda}}\big)}\times P(\tilde{s},\lambda)\times\Bigg(\frac{\tilde{\kappa}_N^2}{\tilde{\kappa}_{J/\Psi}^2}\Bigg)^{-2\big(1+\frac{1}{\sqrt{\lambda}}\big)}\nonumber\\
&&\times  \left(\frac{2}{j_0+\Delta(j_0)}\right)^2\times\mathcal{I}^2(j_0,Q,M_{J/\Psi})\times\frac{\tilde{\kappa}_N^8}{\tilde{\kappa}_{J/\Psi}^8}\times\frac{Q^2}{M_{J/\Psi}^2}
\nonumber\\
&&\times\frac{1}{s^4}\tilde{F}(s)\times 8\times\mathcal{A}^2(j_0,\tau,\Delta, K)\nonumber\\\,,
\eea
where (after noting that $\epsilon_{L}^{\mu}(s\rightarrow\infty)=\frac{q^\mu}{Q}$, $\epsilon_{L}'^{\mu}(s\rightarrow\infty)=\frac{q'^\mu}{M_V}$, and $\theta(s\rightarrow\infty)=0$ hence $\epsilon_{T}\cdot\epsilon'_{T}(s\rightarrow\infty)=-1$, $\epsilon_{T}\cdot A(s\rightarrow\infty)=\epsilon'_{T}\cdot A(s\rightarrow\infty)=0$ for $A=q,q',p,p_1,p_2$)

\be
\tilde{F}(s)=&&\tilde{F}^{TT}(s\rightarrow\infty,t,Q,M_{J/\Psi},m_N)\nonumber\\
=&&9\times\frac{M_{J/\Psi}^2}{Q^2}\times\tilde{F}^{LL}(s\rightarrow\infty,t,Q,M_{J/\Psi},m_N)\nonumber\\
=&&Q^2M_{J/\Psi}^2\times\Big[B_0^{LL}(s\rightarrow\infty,t,Q,M_{J/\Psi})\Big]^2\nonumber\\
=&&\Big[B_1^{TT}(s\rightarrow\infty,t,Q,M_{J/\Psi})\Big]^2\nonumber\\
=&&\lim_{s \to \infty}(q\cdot p\,q'\cdot p)^2=\lim_{s \to \infty}(2q_z^2)^4=\frac{1}{16}\times s^4\,,
\ee
and we have defined the dimensionless function
\bea \label{P}
P(\tilde{s},\lambda)&\equiv & \left[\lambda/\pi^2+ 1\right] ( \sqrt{\lambda}/ 2 \pi )\; \tilde{\xi}^2  \; \frac{e^ {-2\sqrt\lambda  \tilde{\xi}^2 / 2\tilde{\tau}}}{\tilde{\tau}^{3}}\left(1 + {\cal O}\bigg(\frac{\sqrt{\lambda}}{\tilde{\tau}}\bigg) \right)\,,
\eea
with $\tilde{\tau}\equiv\log\tilde{s}=\log[s/\tilde{\kappa}_N^2]$. We have also approximated $(2K^2+8m_N^2)/m_N^2\approx 8$ for small momentum transfer.

\end{widetext}

 \vfil
\end{document}